\documentclass[prb,twocolumn,altaffilletter,nolongbibliography,numerical,flushbottom,secnumarabic,sgroupedaddress,floatfix]{revtex4-2}

\usepackage[polish, english]{babel} % Load Polish language support
\usepackage[T1]{fontenc} % Ensure proper font encoding for Polish characters
\usepackage[utf8]{inputenc} % Use UTF-8 encoding for input
\usepackage{braket}
\usepackage{amsmath}
\usepackage{graphicx}
\usepackage[colorlinks=true, allcolors=blue]{hyperref}
\usepackage{bm}

\usepackage{subcaption}

\usepackage{orcidlink}

\usepackage{bm}

\newcommand{\bfS}{\mathbf{S}}
\newcommand{\bfI}{{\bf I}}
\newcommand{\bfK}{{\bf K}}
\newcommand{\bfT}{{\bf T}}

\newcommand{\up}{\uparrow}
\newcommand{\dw}{\downarrow}

\begin{document}

\title{Exact dimer ground states of long-range spin chains and ladders}

\author{J{{\k e}}drzej Wardyn
\orcidlink{0009-0004-0940-2409}}
\affiliation{Faculty of Physics, University of Warsaw, Pasteura 5, 02-093 Warsaw, Poland}
\email{jedrzej.wardyn@fuw.edu.pl}
%\email{jedrzejwardyn@gmail.com}

\author{Mi{\l}osz Panfil\,\orcidlink{0000-0003-1525-4700}}
\affiliation{Faculty of Physics, University of Warsaw, Pasteura 5, 02-093 Warsaw, Poland}

\begin{abstract}
Interacting spin chains and ladders are known to support a plethora of quantum phases with complex ground-state phase diagrams. In this work, we study a large family of such models and determine precise, explicit conditions under which an exact dimer state is guaranteed to be the ground state. These general conditions are validated for various generalizations of the Majumdar-Ghosh model using exact diagonalization. Our results provide exact reference points in the phase diagrams of a wide class of spin chains and ladders, including those with anisotropic and arbitrary-range interactions.
\end{abstract}

\maketitle

\section{Introduction}

Low-dimensional quantum spin systems can host conventional ordered phases, such as ferromagnetic or antiferromagnetic states, but it has long been known that more exotic orders are also possible. Examples include incommensurate spiral states~\cite{Pati1996ACS}, regular Ne\'el-ordered states for non-frustrated long-range interactions~\cite{Yang2021}, as well as spin-nematic, vector, and scalar chiral phases~\cite{Hikihara2008}. Generally, the emergence of these phases requires interactions beyond nearest neighbors, anisotropy, and/or a ladder geometry. A canonical example is the Majumdar-Ghosh model~\cite{10.1063/1.1664978,Majumdar_1970}, whose ground state is an exact dimer state composed of a product of singlets, also termed valence bond states. This model features antiferromagnetic interactions between both nearest and next-nearest neighbors.

The Majumdar-Ghosh model is quite specific because while it is strongly interacting and nonintegrable, its exact ground state, also in finite chains, is actually known exactly. This property is shared by number of its generalizations~\cite{caspers1982exact,Takano1994,nakano1995exact}, including the spin-$1$ AKLT model \cite{Affleck1987,affleck1988valence} and the linear exchange models~\cite{Kumar2002}. Specifically, the latter belong to the class of models that we study in this work, namely long-range interacting spin-chains.

%\milosz{(I) napisz tutaj o modelach ogolniejsych od MG ale ktore maja spin-1/2 i tylko dwucialowe oddzialywania, 1-2 paragrafy}

The family of long-range interacting spin chains represents a broad class of systems. Rather than focusing heavily on a single model to uncover its specific ground-state phase diagram, our goal is to extract universal, albeit broader, characteristics shared by the entire family. To that end, we consider models defined by Hamiltonians of the form:
\begin{equation} \label{eq:single_site_intro}
    H = \sum_{j=1}^L \sum_{n \geq 1}^{L-1} J_n\, \bfS_{j} \cdot \bfS_{j+n}.
\end{equation}
Here $\bfS_j \cdot \bfS_k = S_j^x S_j^x + S_j^z S_j^z + S_j^z S_j^z$ is the $su(2)$ symmetric coupling between sites $j$ and $k$ with spin-$1/2$ operators $S_j^\alpha$. The coupling constants $J_n$ control strength of the interactions between spins at distance $n$ and can be both  Antiferromagnetic (AFM) $J_n>0$ or Ferromagnetic (FM) $J_n<0$. The competition between FM and AFM interactions of different range has been shown to produce complex phase diagrams containing emergent spin-1 systems with AKLT-like VBS~\cite{Affleck1987,affleck1988valence}, phases in example of FM $J_1$ AFM $J_2$ chain~\cite{agrapidis2019coexistence} and ladder \cite{Agrapidis2017,Wardyn2023}.
Moreover, we do not assume that $J_n$'s are, as would be expected on the physical grounds, decreasing with the distance.

The class of the Hamiltonians described by~\eqref{eq:single_site_intro} includes the nearest neighbor Heisenberg spin chain or the Majumdar-Ghosh model on one end and Calogero-Sutherland systems~\cite{Martin2023} or all-to-all interacting systems~\cite{Ballesteros2025} on the other.

%\milosz{Ten fragment musimy gdzies wplesc w istniejacy tekst:  We also consider Antiferromagnetic (AFM) $J_n>0$ and Ferromagnetic (FM) $J_n<0$ which has been shown to produce complex phase diagrams containing emergent spin-1 systems with AKLT-like VBS~\cite{Affleck1987,affleck1988valence}, phases in example of FM $J_1$ AFM $J_2$ chain~\cite{agrapidis2019coexistence} and ladder \cite{Agrapidis2017,Wardyn2023}.}

In this class of models we ask a very specific question: for which values of the coupling constants $\{J_n\}$ their ground state is the exact dimer state? One solution is the Majumdar-Ghosh model with $J_1 = J_2/2$ and other couplings zero. What are the other solutions? In this work we characterize sets of $\{J_n\}$ for which the ground state is guaranteed to be a dimer phase. Our results thus serve as exact reference points in the reach phase diagrams of such long-range interacting spin chains.

The problem of finding a Hamiltonian with desired properties, such as a specific ground state, is well-established. For instance, a similar challenge arises when constructing parent Hamiltonians for Matrix Product States (MPS)~\cite{PerezGarciaetal2007,Asoudeh2007}. Another related approach involves reconstructing the Hamiltonian directly from ground-state correlation data~\cite{Qi2019} or, in a similar spirit, by considering the properties of the local covariance matrix~\cite{Petrovich2024}. Additionally, an approach based on frustration-free systems leverages the property that the ground state of a small subsystem can be propagated to the entire chain~\cite{Sattathetal2016}. For two-site interacting spin-1/2 models with an exact spin-singlet dimer ground state, a construction based on the tree-graph method has also been developed~\cite{Hikihara2026}.

In this work, working directly with Hamiltonians of the form~\eqref{eq:single_site_intro}, we establish conditions on the coupling coefficients $\{J_n\}$ when: a dimer state is the exact eigenstate of the Hamiltonian and when its energy corresponds to the absolute minimum energy. In fact, we consider slightly more general Hamiltonians than those in Eq.~\eqref{eq:single_site_intro} by allowing for bond alternation, and introducing anisotropy to the interactions.

Such generalizations are consistent with minimal models for the spin-Peierls class of materials \cite{peierls1955quantum,Clarke1994}. The associated spin-Peierls transition is observable through structural changes in X-ray diffraction Bragg peaks~\cite{Abel2007} and via nuclear magnetic resonance~\cite{mayaffre2000nmr}. Dimerization can also be tracked as a sharp drop in SQUID-measured magnetic susceptibility~\cite{Pouget2023}, while the energy gap can be obtained from inelastic neutron scattering~\cite{Pouget2023} or NMR~\cite{mayaffre2000nmr}.

%Such generalizations are in-line with minimal models applicable to spin-Peierls class of materials \cite{peierls1955quantum,Clarke1994}. If a compound undergoes a transition from non-alterated bonds to alterated ones forming dimers we call it spin-Peierls transition. Spin-Peierls phase transition is observable as a change in Bragg peaks visible in X-ray diffraction~\cite{Abel2007} as also by nuclear magnetic resonance\cite{mayaffre2000nmr}. Beside that when the system dimerising can also be found as a sharp drop of the magnetic susceptibility measured by SQUID~\cite{Pouget2023}. The direct way of measuring the gap can be performed by the inelastic neutron scattering~\cite{Pouget2023} or nuclear magnetic resonance~\cite{mayaffre2000nmr}. 

Examples of Spin-Peierls materials include TiOCl~\cite{rotundu2018enhancement, zhang2014transformation}, CuGeO$_3$~\cite{ONeal2017, hase1993observation, meibohm2018comparison}, DF$_5$PNN \cite{inagaki2017phase}, (TMTTF)$_2$PF$_6$~\cite{pouget2017inelastic}, (o-Me$_2$TTF)$_2$NO$_3$~\cite{jeannin2018decoupling} MEM(TCNQ)$_2$~\cite{poirier2013charge}. 
There are also materials with known next-to-nearest bond alteration like mineral szenicsite Cu$_3$(MoO$_4$)(OH)$_4$~\cite{Fujisawa2011,Fujii2015,lebernegg2017frustrated}. Most often such materials can be described with a ladder picture, but due to a spatial structure of the inter-chain couplings the anisotropy has to be accounted for if we choose to rely on purely 1D description~\cite{Xu2021,miljak2005anisotropic,wang2013single,kremer1995anisotropic}. For this reason anisotropic $J_1$-$J_2$ chain and in particular MG model has been considered in the XXZ version~\cite{Kanter1989,nomura1993phase,Gerhardt1998} as well as the XYZ version~\cite{Xu2021}.

The building blocks of dimer ground state, the spin singlets, serve as a basic currency in quantum communication protocols and can be used to form magnetic field resistant qubits in combination with unpolarised triplet states \cite{Petta2005}.
The zero total magnetic momentum and rotational invariance makes such states robust against noise \cite{Dalmonte2015,Chowdhury2024,Gra2014}.
This makes it attractive form the viewpoint of quantum computing, which faces challenges from the decoherence and requires increasingly complex error-correction mechanisms as systems scale.

%\milosz{(II) tutaj o materialach opisywanych przez nasze Hamiltoniany i dlaczego znalezienie fazy dimerowej jest wazne, 1-2 paragrafy}

%\milosz{It has been shown that exact dimer states are very robust against decoherence and noise \cite{Dalmonte2015,Chowdhury2024}. Dimer states can be realized in superconducting circuits (3D Transmon qubits), where we can have longer range interactions.
%Beside that Majumdar-Ghosh and Haldane-Shastry models have been realised in trapped-ion quantum simulators \cite{Gra2014}.}

The rest of the work is organized in the following way. In Section~\ref{sec:setup}, we introduce the class of models and formulate the problem. In Section~\ref{sec:conditions}, we find the sufficient conditions on the coupling constants for the dimer state to have the lowest energy. These conditions are then analyzed in Section~\ref{sec:polytopes} and shown to have a geometric interpretation in terms of polytopes in the space of the couplings. In Section~\ref{sec:anisotropic} we generalize the construction to anisotropic interactions. In the following Section~\ref{sec:examples} we illustrate our findings and verify against numerical computations for few generalizations of the Majumdard-Ghosh model. We also consider Calogero-Sutherland model and its generalization, the Inozemtsev chain. The conclusions are presented in Section~\ref{sec:final}. The Appendices~\ref{app:exact_dimers}, \ref{app:LEM} and ~\ref{app:top} contain technical steps regarding the pseudo-spin representation, semi-positivity of the linear exchange models and spectrum of the antisymmetric quantum top, while in Appendix~\ref{app:triplets} we show that our formalism can also be adapted to triplets ground states.

\vspace{-0.25cm}

\section{Long-range spin chains} \label{sec:setup}

% \milosz{do poprawy jak ustalimy wstep}

% In this work, we consider long-range interacting spin chains with Hamiltonians of the very general form
% \begin{equation} \label{single_site}
%     H = \sum_{j=1}^N \sum_{n \geq 1}^{N-1} J_n\, \bfS_{j} \cdot \bfS_{j+n}.
% \end{equation}
% Here $\bfS_j \cdot \bfS_k = S_j^x S_j^x + S_j^z S_j^z + S_j^z S_j^z$ is the $su(2)$ symmetric coupling between sites $j$ and $k$ with spin-$1/2$ operators $S_j^\alpha$. The coupling constants $J_n$ control strength of the interactions between the spins at distance $n$ and at this stage we do not make any assumptions on them. Specifically, we do not assume that $J_n$'s are, as would be expected on the physical grounds, decreasing with the distance. The class of the Hamiltonians described by~\eqref{single_site} is very large and includes the nearest neighbour Heisenber spin chain or Majumdar-Ghosh model on one end and Calogero-Sutherland systems~\cite{Martin2023} or all-to-all interacting systems~\cite{Ballesteros2025} on the other. 

%We consider antiferromangetic couplings, $J_k > 0$ %and assume that there is $k_{\rm max}<N$ such that $J_k = 0$ for $k>k_{\rm max}$.  

\begin{figure}
	\centering
	\includegraphics[width=0.7\columnwidth]{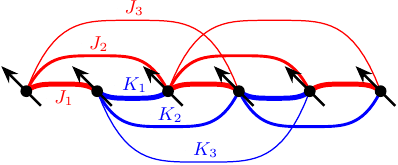}
	\caption{Diagram of a $J_1-K_1-J_2-K_2-J_3-K_3$ chain given in \eqref{eq:double_site}. The $J_n$($K_n$) interactions are drawn in red (blue) and the decreasing thickness signifies increase in distance $n$.}
    \label{fig:single_site}
\end{figure}

\begin{figure}
	\centering
	\includegraphics[width=0.55\columnwidth]{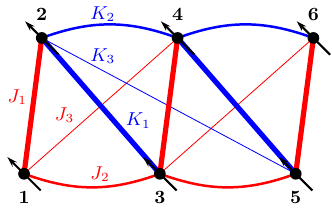}
	\caption{Diagram of a $J_1-K_1-J_2-K_2-J_3-K_3$ ladder given in ~\eqref{eq:double_site} with the same color-thickness convention as in the Fig.~\ref{fig:single_site} to which this ladder diagram is topologically equivalent.
    }
    \label{fig:double_site}
\end{figure}

We consider a family of the long-range interacting spin models with the bond alteration described by the Hamiltonians of the form
\begin{equation} \label{eq:double_site}
    H = \!\sum_{j}^L \sum_{n \geq 1}^{L-1}\!\left( J_n\, \mathbf{S}_{2j-1} \cdot \mathbf{S}_{2j-1+n} + K_n\, \mathbf{S}_{2j} \cdot \mathbf{S}_{2j+n}\right).
\end{equation}
They can be seen as generalizations of Hamiltonian~\eqref{eq:single_site_intro} from the Introduction to unit cells formed by two consecutive sites with distinct couplings between even and odd sites,
The long-range structure of couplings is shown in Fig.~\ref{fig:single_site}. The doubled unit cell can be also understood as geometry of a ladder as shown in Fig~\ref{fig:double_site}. We assume, without a loss of generality, that $J_1 \geq K_1$. Therefore, if dimers are formed, they will be situated at the rungs of the ladder. The long-range chain~\eqref{eq:single_site_intro} is a special case of the long range-ladder with $J_n = K_n$. 

In Section~\ref{sec:anisotropic},  we will also consider the anisotropic versions of the two models with the component-dependent interactions between the spins,
\begin{equation}
    J_n \bfS_j \cdot \bfS_{j+n} \rightarrow J_n^x S_j^x S_j^x + J_n^y S_j^y S_j^y + J_n^z S_j^z S_j^z.
\end{equation}

Within the class of Hamiltonians~\eqref{eq:single_site_intro} and~\eqref{eq:double_site} we will look for those whose ground state has the product form
\begin{equation} \label{product_state}
    |{\rm gs} \rangle = \bigotimes_{i}^{L/2} |i, i+1 \rangle,
\end{equation}
with the building block $|i, i+i \rangle$ involving the neighboring sites. For definiteness, when writing~\eqref{product_state} we assumed that the chain is of even length. Throughout the text we will consider systems with either open (OBC) or periodic (PBC) boundary conditions. However, the chains will be always of even number of sites $L$ to avoid extra complications due to an unpaired spin.

The elementary blocks $|i, i+1 \rangle$ appearing in~\eqref{product_state} are formed by the neighboring sites and can be of different nature. Studied cases range from simple structures in ferromagnetic or antiferromagnetic chains to more intricate triplet dimer or nematic states~\cite{agrapidis2019coexistence,Pati1996ACS,Yang2021,Hikihara2008}. Our sole focus here is on the singlet dimer
\begin{equation} \label{dimer}
    |d_{i, i+1}\rangle = |s\rangle \equiv \frac{1}{\sqrt{2}}\left(\ket{\up\dw}-\ket{\dw\up} \right).
\end{equation} 
We note that there are also known cases where the elementary blocks are either larger—composed of three or more sites, or consist of sites spaced further apart, as in Anderson's RVB state~\cite{Baskaran1987}. Such cases, however, fall beyond the scope of the present work.

The aim of this work is to provide exact characterization of the regions in the ground-state phase diagrams of interacting spin chains and ladders with the exact dimer order. The question we ask is: for what values of $\{J_n\}$ or $(\{J_n\},\{K_n\})$ the exact dimer state
\begin{equation} \label{dimer_state}
    |D\rangle = \bigotimes_{i}^{L/2} |d_{i, i+1} \rangle,
\end{equation}
with $|d_{i, i+1} \rangle$ given in~\eqref{dimer},
is the ground state of the Hamiltonians~\eqref{eq:single_site_intro} and~\eqref{eq:double_site} respectively?

Partial results in this direction exists, for example, in the linear exchange models~\cite{Kumar2002} the ground state is the dimer when $J_n = K_n$ and with maximal range of interactions $n_{\rm max}$ being even. The coupling constants are then fixed to be 
\begin{equation} \label{LEM_condition}
    J_n = \frac{n_{\rm max} - n + 1}{n_{\rm max}}, \qquad n = 1, \dots, n_{\rm max},
\end{equation}
and $J_n = 0$ for $n > n_{\rm max}$. We will find that the dimer phase actually exists in a much larger region of the phase space (also without the bond alteration) then the above condition predicts.  

The class of the Hamiltonians considered is indeed reach and as already mentioned contains canonical models such like the XXX Heisenberg chain (where only $J_1$ is nonzero) or Majumdar-Ghosh model (with $J_1 = 2 J_2$ and all other $J_n$ zero). These two models have very different ground state. For the antiferromagnetic Heisenberg chain it is a highly entangled, critical state with logarithmic scaling of the entanglement entropy. Instead, for the Majumdar-Ghosh model it is a product state exactly of the form~\eqref{dimer_state}. In the general case for systems~\eqref{eq:single_site_intro} and \eqref{eq:double_site} their ground state structure is not known.

Before getting into the details, let us summarize our main findings. To formulate the conditions on the coupling constants we introduce the following combinations 
\begin{equation} \label{eq:alpha_beta_J_Jp_setup}
    \begin{aligned}
        \alpha_n &= J_n - J_{n+1} - K_{n+1} + K_{n+2}, \\
        \beta_n &= K_n - K_{n+1} - J_{n+1} + J_{n+2},
    \end{aligned}
\end{equation}
The sufficient conditions for the exact dimer ground state are then
\begin{equation}
    \alpha_n \geq 0, \qquad \beta_{2n} \geq 0, \qquad \beta_{2n+1} = 0.
\end{equation}
In the anisotropic case, the conditions are more complicated and here we formulate them only for anisotropy in a single component. 
The conditions are then
\begin{equation}
    \begin{aligned}
    &\alpha_n^z \geq - \frac{2}{n+2} \alpha_n, \quad \beta_{2n}^z \geq - \frac{2}{n+2} \beta_{2n}, \\
    &\beta_{2n+1}^z = \beta_{2n+1} = 0 \vphantom{\frac{1}{2}},
    \end{aligned}
\end{equation}
where $\alpha_n^z$ and $\beta_n^z$ are the combinations of the coupling coefficients $J_n^z$ and $K_n^z$ standing now in front of the $s_{i}^z s_{i+n}^z$ terms in the Hamiltonian. The $x$- component and $y$-component parts of the Hamiltonian are not modified.

\section{Exact dimer ground states} \label{sec:conditions}

Let us recall the precise question we are asking: under what conditions on the coupling constants $\{J_n\}$ or $\{J_n, K_n\}$, do the Hamiltonians~\eqref{eq:single_site_intro}~and~\eqref{eq:double_site} have the exact dimer state $|D\rangle$ of eq.~\eqref{dimer_state} as their ground state?
%We answer this question in two steps: first, we find the conditions on the coupling constants and second, we interpret them in the space of models defined by~\eqref{single_site} and~\eqref{eq:double_site}.

For a state $|D\rangle$ to be a ground state of a Hamiltonian $H$, it has to be an eigenstate with the smallest energy. Remember, that in our problem $D$ is fixed and we are looking for conditions on the Hamiltonian. How to find these conditions in practice?  The approach that we are going to use is the following. First we find the Hamiltonians for which $|D\rangle$ is an eigenstate. This  gives as a first set of constraints on the coupling constants. In the process, we will also determine its eigenenergy $E_D$: $H|D\rangle = E_D|D\rangle$. In the second step we consider Hamiltonian $\bar{H} = H - E_D$. By construction it annihilates the dimer state, $\bar{H}|D\rangle = 0$. Therefore, if we can prove that $\bar{H}$ is semi-positive definite then we prove that the dimer state is the ground state of $H$. The semi-positivity of $\bar{H}$ introduces a second set of constraints on the coupling constants.

Few comments are in place. First, this procedure is exact and does not rely on thermodynamic limit. All the statements will be true in a finite system. Secondly, whereas any bounded operator can be written as some semi-positive operator plus a constant term, it might be difficult in practice to find such representation. We will provide explicit solution to this problem for the class of the long-range Hamiltonians~\eqref{eq:double_site}.

Thirdly, 
%the above construction generalizes to Hamiltonians that can be written as sums of two (or more) semi-positive definite operators. If all these operators have exact dimer ground states, their linear combination will inherit it.
%Finally, 
this approach tell us that if certain conditions are fulfilled, then the ground state is the dimer state. It tells us nothing about the situation when the conditions are not fulfilled. Specifically, it does not rule out the dimer ground state for other values of the coupling coefficients. %However, as we shall see, from numerical computations and for isotropic models the conditions are if and only if. The situations complicates for the anisotropic models and we leave this discussion for Section~\ref{sec:anisotropic}. 
In that respect it is instructive to find a necessary condition for the dimer phase. Then, if such condition is broken we are guaranteed that the ground state is not the exact dimer state. This condition will provide a useful reference point while charting the ground state phase diagrams. 

For the dimer ground state, its energy should be lower then the energy of any other eigenstate. The convenient state to compare its energy with is the ferromagnetic state. This is clearly an eigenstate of any of the Hamiltonians considered and its energy is 
\begin{equation}
    E_{\rm FM} = \frac{L}{8}\sum_{n \geq 1}\left(J_n + K_n\right).
\end{equation}
Therefore, the necessary condition for the dimer ground state is $E_{\rm FM} \geq E_{\rm D}$ or equivalently
\begin{equation}
    %\sum_{n \geq 1}\left(J_n + K_n \right) \geq - 3 J_1,
    %J_1 \geq - \frac{1}{3} \sum_{n \geq 1}\left( J_n + K_n \right).
    J_1 \geq - \frac{1}{4}K_1 - \frac{1}{4} \sum_{n \geq 2}\left( J_n + K_n \right),
    \label{eq:FM_transition_condition}
\end{equation}
where we have used the fact that $E_D = - 3/8\, J_1 L$ which we will calculate in the next section. For positive $J_1$, favoring the antiferromagnetic phase, this condition reveals how strongly ferromagnetic the remaining interactions are required, for the ferromagnetic order to take over the dimer order.  

After this quick detour, we come back to the question of the sufficient condition. 
We start, in the next section, by determining the conditions on the coupling constants for the dimer state to be an exact eigenstate and later answer the question of the semi-positivity.

\subsection{Exact dimer states}

We consider the more general case of the long-range ladder and find conditions on the coupling constants for the dimer state to be an exact eigenstate and also determine its eigenenergy. 

Following~\cite{Lamas2015} we introduce the pseudospin operators
\begin{align} \label{pseudospin}
{\bf I}_{i}= {\bfS}_{2i}+ \bfS_{2i-1} \quad\quad {\bfT}_{i} = \bfS_{2i}- {\bfS}_{2i-1},
\end{align}
forming a Lie algebra with single-site commutation relations ($\mu, \nu = x,y,z$)
\begin{equation}
\label{eq:localliealgebra}
\begin{aligned}
   \left[I_{\mu},I_{\nu}\right]&= {\bf i} \epsilon_{\mu\nu\eta} I_{\eta}\\
   \left[I_{\mu}, T_{\nu}\right]&= {\bf i} \epsilon_{\mu\nu\eta} T_{\eta}\\
   \left[T_{\mu}, T_{\nu}\right]&= {\bf i} \epsilon_{\mu\nu\eta} I_{\eta}
\end{aligned}
\end{equation}
where $\epsilon_{\mu\nu\eta}$ is the fully antisymmetric Levi-Civita symbol. The pseudospins of different sites commute with each other. The usefulness of the pseudospin representation stems from the fact that ${\bf I}$ operator annihilates the singlet state, ${\bf I}_j |d_{j, j+1}\rangle = 0$.

In terms of the pseudospin operators the long-range ladder Hamiltonian~\eqref{eq:double_site} is 
% \begin{equation} \label{H_PS_plus_identity}
%     %H = H_{\rm PS} + J_1 \sum_{j=1}^N \bfS_j^2,
%     H = H_{\rm PS} + \frac{1}{2}\left(J_1 + J_1'\right)\mathcal{E}_0  , \qquad \mathcal{E}_0 = \frac{3}{4}N,
% \end{equation}
% or
\begin{equation} \label{H_PS_plus_Heis}
    %H = H_{\rm PS} + J_1 \sum_{j=1}^N \bfS_j^2,
    H = H_{\rm PS} + J_1 \sum_{j=1}^{L/2} \bfS_{2j-1} \cdot \bfS_{2j},
\end{equation}
with
\begin{align} 
    H_{\rm PS} = \frac{1}{4} \sum_{j=1}^{L/2} \sum_{n\geq 1} \left( J_n^{II} \bfI_j \cdot \bfI_{i+n} + J_n^{TT} \bfT_j \cdot \bfT_{j+n} \right. \nonumber \\
    \left. + J_n^{IT} \bfI_j \cdot \bfT_{j+n} + J_n^{TI} \bfT_j \cdot \bfI_{j+n}  \right). \label{H_PS}
\end{align} 
The calculations are straightforward and are presented in Appendix~\ref{app:exact_dimers}.
The coupling constants of the pseudospin Hamiltonian $H_{\rm PS}$ are
\begin{equation}
\begin{aligned}
    J_n^{II} &= K_{2n} + J_{2n} + K_{2n-1} + J_{2n+1}, \\
    J_n^{IT} &= K_{2n} - J_{2n} - K_{2n-1} + J_{2n+1}, \\
    J_n^{TI} &= K_{2n} - J_{2n} + K_{2n-1} - J_{2n+1}, \\
    J_n^{TT} &= K_{2n} + J_{2n} - K_{2n-1} - J_{2n+1}.
    \label{eq:dimerisation_pseudospin}
\end{aligned}
\end{equation}
Since operators $\bfI_i$ annihilate the dimer state and operators $\bfI_j$ and $\bfT_k$ commute for $j\neq k$, the sufficient condition for the dimer state to be an eigenstate of $H_{\rm PS}$ is 
%, while $\bfT$ annihilates the triplet state $|t\rangle = \frac{1}{\sqrt{2}}\left(\ket{\up\dw}-\ket{\dw\up} \right)$.
that $J_n^{TT}=0$ or, returning to the original coupling constants,
\begin{equation} \label{condition_dimer}
    K_{2n} + J_{2n} = K_{2n-1} + J_{2n+1}.
\end{equation}
In fact, once this condition is fulfilled, the dimer state is annihilated by $H_{\rm PS}$. 
%The other term in $H$, related to the Casimir operator (see Apppendix~\ref{app:exact_dimers}),  just shifts  the whole spectrum. Redefining the original Hamiltonian to compensate for this shift, we conclude that the dimer state is annihilated by the long-range ladder if the coupling constants obey~\eqref{condition_dimer}.

The dimer state formed along the rungs of the ladder is also an eigenstate of the Heisenberg-like term in~\eqref{H_PS_plus_Heis} with the energy $- J_1 \mathcal{E}_0$ where we defined $\mathcal{E}_0 = 3/8 L$. This implies that once the condition~\eqref{condition_dimer} is fulfilled, the dimer is an exact eigenstate of the long-range ladder~\eqref{eq:double_site} with energy $E_D = -J_1 \mathcal{E}_0$. 

Therefore, following the procedure outlined above, we define
\begin{equation}
    \bar{H} = H - E_D, {\text{ such that} }\; \bar{H} |D\rangle = 0,
\end{equation}
with the coupling constants constrained by~\eqref{condition_dimer}.
This result completes the first step in determining the conditions for the exact dimer phase.

\subsection{Conditions for the exact dimer ground state}

The second step consists of determining the conditions under which the (shifted) long-range Hamiltonians $\bar{H}$ are semi-positive definite. 
 %It is enough to verify this for the coupling constants obeying~\eqref{condition_dimer}.

The strategy is to define semi-positive operators that will serve as building blocks for the whole Hamiltonian. The resulting Hamiltonian is guaranteed to be semi-positive definite, if it is a linear combination of these semi-positive operators with positive coefficients.
The building blocks are the linear exchange models (LEM) defined for the two legs of the ladder as follows
\begin{equation} \label{LEM_def}
    \begin{aligned}
    H_n^{1} &= \frac{1}{2}\sum_{j=1}^{L/2} \left(\bfS_{2j-1} + \cdots + \bfS_{2j+n-1} \right)^2 - E_{n, 0}, \\
    H_n^{2} &= \frac{1}{2} \sum_{j=1}^{L/2} \left(\bfS_{2j} + \cdots + \bfS_{2j+n} \right)^2 - E_{n, 0}.
    \end{aligned}
\end{equation}
Index $n$ denotes the maximal range of interactions present in $H_{n}^{1,2}$ and 
\begin{equation}
    E_{n, 0} = \begin{cases}
        \quad 0 , \qquad n \text{ odd}, \\
        \frac{1}{2}\mathcal{E}_0, \qquad n \text{ even}.
    \end{cases}
\end{equation}
The LEM Hamiltonians are semi-positive definite and the energy shift $E_{n,0}$ is chosen such that their lowest energy state for even $n$ has exactly zero energy, see Appendix~\ref{app:LEM} for the details. For even $n$, the ground state of $H_{n}^1 + H_n^2$ is known to be the dimer state~\cite{Kumar2002}, while the combination $H_1^1 + H_2^1$ is the integrable XXX model. Thus, the family of the linear exchange models encompasses the variety of physics expected in the 1d spin models. 

In fact, the LEM Hamiltonians, supplemented with the identity operator, form an operatorial basis for any long-range ladder Hamiltonian of the form~\eqref{eq:double_site}. Therefore, the ladder Hamiltonian can be expressed as
\begin{equation} \label{linear_LEMs}
    H = \sum_{n=1} \left(\alpha_n H_n^1 + \beta_n H_n^2 \right) - E_0,
\end{equation}
with suitably chosen coefficients $\{\alpha_n\}$ and $\{\beta_n\}$. The energy shift
\begin{align} \label{LEM_shift}
    E_0 &= \frac{1}{2} \mathcal{E}_0 \left[ \sum_{n\geq 1} n\left(\alpha_n + \beta_n\right) + \alpha_{2n-1} + \beta_{2n-1} \right],
\end{align}
takes into account $E_{n, 0}$ and the contributions from the single site Casimir $\bfS_i^2$ operators which appear in the LEM's but are absent in the long-range ladder Hamiltonian~\eqref{eq:double_site}. 
%Since we are working with periodic boundary conditions, it is sufficient to consider LEM's with $n_{\rm max} \leq  L/2$. They cover all possible interaction ranges in a system of length $L$ with periodic boundary conditions. 

The Hamiltonian~\eqref{linear_LEMs} for $\alpha_n \geq 0$ and $\beta_n \geq 0$ has spectrum bounded from below by $E_0$. Therefore, what remains to be done is to relate the coefficients $\alpha_n$ and $\beta_n$ to the coupling constants $\{J_n\}$ and $\{K_n\}$. 

%The relation between coefficients $\alpha_n$, $\beta_n$ and the coupling constants $J_n$ and $K_n$ can be established with direct computations. We distinguish two cases: finite interactions range with $n_{\rm max}/L$ vanishing in the thermodynamic limit and the complementary case with the interaction range being a finite fraction of the system size.

\subsection{Solution for the coupling constants}

In the following we assume that there is no coupling at distances larger then $L/2$. In the system with periodic boundary conditions this does not introduce any restriction on the possible interactions. In this situation, sums over $n$ in~\eqref{linear_LEMs} and~\eqref{LEM_shift} truncate at some $n_{\rm max} \leq L/2$. We then find the following relations between the coefficients $\alpha_n$ and $\beta_n$ of the LEM's and the coupling constants $J_n$, $K_n$,
\begin{equation}
    \begin{aligned}
        J_n = \sum_{j=n}^{n_{\rm max}} \lfloor \frac{j+1}{2}\rfloor \alpha_j + \sum_{j=n+1}^{n_{\rm max}} \lfloor\frac{j}{2}\rfloor \beta_j, \\
        K_n = \sum_{j=n}^{n_{\rm max}} \lfloor \frac{j+1}{2}\rfloor \beta_j + \sum_{j=n+1}^{n_{\rm max}} \lfloor\frac{j}{2}\rfloor \alpha_j, 
    \end{aligned}
\end{equation}
where $\lfloor x \rfloor$ is the floor function returning the largest integer smaller then $x$.
These relations can be inverted. The result is
\begin{equation} \label{eq:alpha_beta_J_Jp}
    \begin{aligned}
        \alpha_n &= J_n - J_{n+1} - K_{n+1} + K_{n+2}, \\
        \beta_n &= K_n - K_{n+1} - J_{n+1} + J_{n+2},
    \end{aligned}
\end{equation}
with the convention that $J_n = K_n = 0$ for $n > n_{\rm max}$.
These relations express the long-range ladder model as a sum of semi-positive definite operators. To conclude that the Hamiltonian of the long-range ladder is semi-positive definite requires that the coefficients $\alpha_n$ and $\beta_n$ are nonnegative. We will analyze the implications of this condition on the coupling constants $J_n$ and $K_n$ in the next section. %At this stage let us assume that for some nonnegative $\alpha_n$ and $\beta_n$ we in fact obtain long-range ladder Hamiltonians with nonnegative coupling constants.   

The dimer eigenstate condition~\eqref{condition_dimer} puts further constraints on the parameters of the linear combination~\eqref{linear_LEMs}. Comparing the condition~\eqref{condition_dimer} with relations~\eqref{eq:alpha_beta_J_Jp}, we conclude that it is equivalent to additionally imposing $\beta_{2n-1} = 0$. 

Furthermore, it is a matter of straightforward computations to find that in such case the energy shift $E_0$ is exactly equal to $E_D$. Therefore, the linear combinations of LEM's with $\beta_{2n-1}=0$ exactly realizes the $\bar{H}$ Hamiltonian,
\begin{equation}
    \bar{H} = \sum_{n}^{n_{\rm max}} \left(\alpha_n H_n^1 + \beta_n H_n^2 \right), \qquad \beta_{2n-1} = 0,
\end{equation}
with the coefficients related by~\eqref{eq:alpha_beta_J_Jp}. 

In summary, we have found that long-range ladder Hamiltonian~\eqref{eq:double_site} has the energy bounded from below by $E_0$, see eq.~\eqref{LEM_shift}, when coefficients $\alpha_n$ and $\beta_n$ determined from the coupling coefficients $J_n$ and $K_n$ through relations~\eqref{eq:alpha_beta_J_Jp} are nonnegative. We shall refer to them as semi-positivity conditions as these are conditions for the Hamiltonian $H - E_0$ to be semi-positive definite,
\begin{equation}
    \alpha_n \geq 0,\; \beta_n \geq 0, \qquad {\text{semi-positivity}}.
\end{equation}

Furthermore, if, in addition to other coefficients being nonnegative, coefficients $\beta_{2n-1} =0$, then the long-range ladder Hamiltonian~\eqref{eq:double_site} has the exact dimer ground state with the energy $E_D$
\begin{equation}
    \alpha_n \geq 0,\; \beta_{2n} \geq 0,\; \beta_{2n-1} = 0 \qquad {\text{dimer phase}}
\end{equation}

In Section~\ref{sec:polytopes}, we will analyze the implications of these conditions, here stated for $\alpha_n$ and $\beta_n$, on the coupling coefficients $J_n$ and $K_n$. %For now, we turn to the cases of all-to-all interactions.  

Before doing so, let us specialize to the chain case. Here, the situation is quite straightforward, for $J_k = K_n$ implies $\alpha_n = \beta_n$ and the conditions for the dimer phase are $\alpha_{2n-1} = 0$ and $\alpha_{2n} \geq 0$. In terms of the coupling constants $J_n$ we have the self-averaging and convexity properties,
\begin{equation}
    \begin{aligned}
        &J_{2n} = \frac{1}{2} \left( J_{2n-1} + J_{2n+1}\right), \\
        &J_{2n-1} - 2 J_{2n+1} + J_{2n+3} \geq 0.        
    \end{aligned}
    \label{eq:dimerisation_conditions_isotropic}
\end{equation}
The first relation implies that the largest non-zero coefficient must be even for the exact dimer ground state. For example, model with $J_1$ only is the Heisenberg nearest neighbors XXX spin chain which ground state not even of the product form. However, adding one lattice longer range interaction with $J_2 = J_1/2$ leads to the Majumdar-Ghosh model with the exact dimer ground state. Adding $J_3$ breaks again the product state structure of the ground state which is yet again restored by admixture of $J_4$ term and so on. In the simplest case this leads back to the structure of a single LEM and conditions~\eqref{LEM_condition}, however also other solutions are possible. We will see an illustration of this in Section~\ref{sec:examples}.

% This shows that for non-negative coefficients $\alpha_n$ and $\beta_n$ the energy of the long-range ladder is bounded from below by $E_0$. Earlier we have shown that the dimer state has the energy $E_0$. This shows that the dimer state is the eigenstate.

% In summary we have shown that the long-range ladder with the coupling coefficients $J_n$ and $K_n$ such that $\alpha_n$ and $\beta_n$ given in~\eqref{eq:alpha_beta_J_Jp} are nonnegative, then the Hamiltonian is semi-positive definite. If additionally, $\beta_{2n-1} = 0$, then its ground state is the dimer state. 

% when the following conditions are fulfilled
% \begin{equation}
%     \begin{aligned}
%         J_k - J_{k+1} - J_{k+1}' + J_{k+2}' \geq 0 \\
%         J_{2k}' - J'_{2k+1} - J_{2k+1} + J_{2k+2} \geq 0, \\
%         J_{2k}' + J_{2k} = J_{2k-1}' + J_{2k+1}.
%     \end{aligned}
% \end{equation}

\section{Space of models} \label{sec:polytopes}

We will analyze now the conditions on the coupling coefficients $J_n$ and $K_n$ for semi-positivity of the Hamiltonian and for the dimer phase. The relevant mathematical structures are these of a convex cone and of polytopes. The main aim of this section is thus to reinterpret the conditions on semi-positivity and dimer phase geometrically. We begin by recalling the notion of the convex cone and related concepts.

A {\em convex cone} is defined as a set of objects closed under linear combinations with nonnegative coefficients: set $C$ is a convex cone if for any $a,b \in C$, $\alpha a + \beta b \in C$ for $\alpha, \beta \geq 0$. There is also a notion of a convex subcone. Namely, $D$ is a {\em convex subcone} of $C$ if $D$ is a convex cone and $D \subset C$. Finally, the last ingredient we need is the face $F$. {\em Face} $F$ is a convex subcone of $C$ such that for any $a,b \in C$ such that $a+b \in F$ then $a,b \in F$. We apply now these concepts to the space of spin Hamiltonians. 
%This is always a finite number, since as stated in the introduction, we assume that there is a $n_{\rm max}$ such that $J_n = K_n = 0$ for $n > n_{\rm max}$.  

The first observation is that the space of models hosting the exact dimer ground state takes the form of a convex cone. This follows from the two following observations. First one is that semi-positive Hamiltonians form a convex cone $C_{\rm pos}$. We have already used this property when considering a linear combination of LEM's and requiring the coefficients to be nonnegative, $\alpha_n, \beta_n \geq 0$. The resulting Hamiltonians are semi-positive, thus semi-positive Hamiltonians form a convex cone.

Furthermore, the dimer condition $\beta_{2n-1} = 0$ implies that the space $C_{\rm dimer}$ of long-range ladders hosting the exact dimer ground state is also a convex cone. This is because any nonnegative linear combination of Hamiltonians having dimer ground state has the dimer ground state. The two cones are of course related, $C_{\rm dimer} \subset C_{\rm pos}$ hence $C_{\rm dimer}$ is a convex subcone of $C_{\rm pos}$. 

Finally, from the dimer condition it is immediate that $C_{\rm dimer}$ is a actually a face of $C_{\rm pos}$: if a convex sum of semi-positive Hamiltonians has the dimer ground state, each Hamiltonian individually has the dimer ground state. 

This establishes a relation between the spaces of semi-positive and dimerized Hamiltonians on one hand and the structure of convex cone on the other hand. We will now use a relationshp between convex cones and polytopes to provide geometrical illustrations for the two spaces. 

Convex cones are related to polytopes. A polytope is a set of points in the $n$-dimensional space that has flat sides. We will define it as an intersection of finite number of half-spaces defined by $\alpha_n \geq 0$ and $\beta_n \geq 0$. This constructs the polytope related to a semi-positive Hamiltonian. By further setting $\beta_{2n-1}=0$ chooses a face of this polytope which describes then the sought-after family of Hamiltonians with dimer ground state.

The dimensionality of the problem is controlled by twice the number of non-zero coupling coefficients. For example, if we consider the most general ladder Hamiltonian with interactions of maximal range $3$, we denote it $J_1$--$J_2$--$J_3$, the phase space of such models is $6$ dimensional. The semi-positivity conditions define a $6$-dimensional polytope. The dimerization condition is a face of this polytope defined by conditions $\beta_1 = \beta_3 = 0$, thus it is $4$-dimensional polytope. 

To easier visualize these polytopes and to expose the role of the ladder geometry, namely the possibility that $J_n \neq K_n$, we introduce average and difference of the coupling coefficients of a given range,
\begin{equation}
    \bar{J}_n = \frac{1}{2}\left(J_n + K_n\right), \qquad \delta_n = J_n - K_n.
\end{equation}
We will refer to $\delta_n$ as the rung-leg coefficients: for $\delta_n > 0$, the dimers are more likely to be formed on the rungs of the ladder. The expressions for $\alpha_n$ and $\beta_n$ in terms of these new parameters are
\begin{equation}
    \begin{aligned}
        \alpha_n &= \bar{J}_n -2 \bar{J}_{n+1} + \bar{J}_{n+2} + \frac{1}{2}\left( \delta_n - \delta_{n+2}
\right), \\
        \beta_n &= \bar{J}_n -2 \bar{J}_{n+1} + \bar{J}_{n+2} - \frac{1}{2}\left( \delta_n - \delta_{n+2}
\right)
    \end{aligned}
\end{equation}
The semi-positivity conditions $\alpha_n \geq 0$ and $\beta_n \geq 0$ are equivalent to
\begin{equation} \label{half-spaces}
    \bar{J}_n - 2 \bar{J}_{n+1} + \bar{J}_{n+2} \geq \frac{1}{2}|\delta_{n} - \delta_{n+2}|.
\end{equation}
This condition implies that the averaged coefficients $\bar{J}_n$ has to form a convex sequence with the second difference bounded from below by the difference of the rung-leg coefficients. For fixed values of rung-leg coefficients $\delta_n$ each condition defines a half-space in the space of $\{\bar{J}_n\}$. All conditions together define again a convex polytope, now in the space of dimension equal to the number of non-zero coupling coefficients (and not double of it as in the original variables $J_n$ and $K_n$).

The dimerization condition, $\beta_{2n-1}=0$, reads now
\begin{equation} \label{dimer_condition}
    \bar{J}_{2n-1} - 2 \bar{J}_{2n} + \bar{J}_{2n+1} = \frac{1}{2}(\delta_{2n-1} - \delta_{2n+1}).
\end{equation}
We notice that due to relation~\eqref{half-spaces}, the dimerization condition can only be fulfilled if the odd rung-leg coefficients form a non-increasing sequence: $\delta_{2n-1} \geq \delta_{2n+1}$.

\begin{figure}
    \includegraphics[scale=0.5]{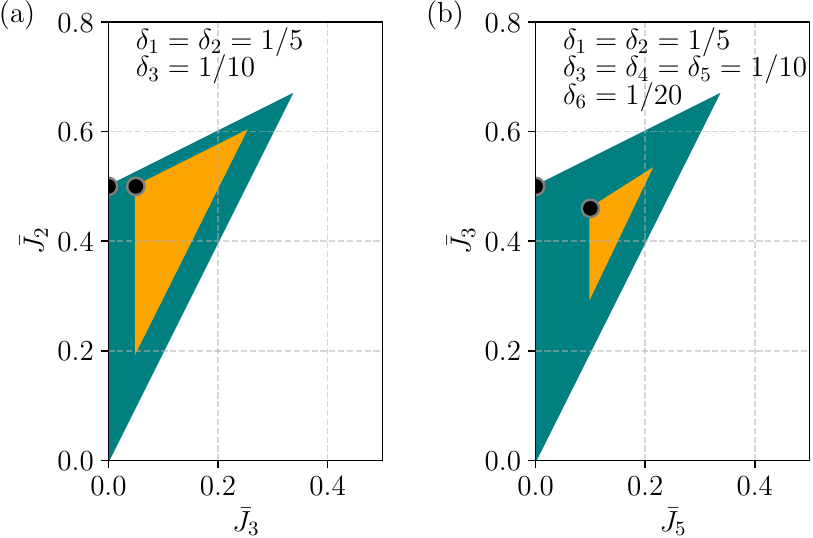}
    \caption{(a) The positivity polytope for $J_1$--$J_2$--$J_3$ model: the teal region is for the spin chain ($\delta_n=0$) and the orange region for ladder with $\delta_k$ specified on the plot. The values are in the units of $J_1$. The two black dots are the models with the exact dimer ground states. For the chain this is the Majumdar-Ghosh model. (b) The dimerization polytope for $J_1-J_5$ model for chain (teal) and for ladder (yellow). The values of $\bar{J}_2$ and $\bar{J}_4$ are fixed by the condition~\eqref{dimer_condition}.}
    \label{fig:polyplot}
\end{figure}

Examples of the polytopes defining the semi-positive and dimerized Hamiltonians $J_1$--$J_2$--$J_3$ are shown in Fig.~\ref{fig:polyplot}. 
The revealed structure has the following properties. The dimerization condition for the ladder is inside the dimerization condition for the chain. This follows immediately from~\eqref{half-spaces} since once the equalities are fulfilled for $\delta_n \neq 0$, then they are also fulfilled for $\delta_n=0$.

%The convex cone structure implies also that models with exact dimer states can be combined into new models sharing the same exact ground state.

%\cite{nakano1997long}

\section{Anisotropic interactions} \label{sec:anisotropic}

We discuss now generalizations of Hamiltonians~\eqref{eq:single_site_intro} and~\eqref{eq:double_site} to anisotropic interactions. Each interaction term in the two Hamiltonians is modified according to
\begin{equation}
    J_n\, \bfS_j \cdot \bfS_{j+n} \longrightarrow J_n^{\mu} S_j^{\mu} S_{j+n}^{\mu},
\end{equation}
with the summation over $\mu = x,y,z$ implied and analogously for the $K$-terms. 

The condition for the dimer state to be an eigenstate of the anistropic Hamiltonians is readily a simple generalization of the condition in the isotropic case. Indeed, the rewriting in terms of the pseudospin operators proceeds in the same way but now component wise. In the result, we find the same type of the condition but for each component separately
\begin{equation} \label{dimerization_condition_ani}
    K_{2n}^{\mu} + J_{2n}^{\mu} = K_{2n-1}^{\mu} + J_{2n+1}^{\mu}, \qquad \mu = x,y,z.
\end{equation}
The energy shift is in this case 
\begin{equation}
    \mathcal{E}_0 = \frac{1}{8}\left( J_1^x + J_1^y + J_1^z \right)N.
\end{equation}
Thus, the dimer state is an eigenstate of $H - \mathcal{E}_0$ of zero energy with $H$ anisotropic version of either the long range spin chain~\eqref{eq:single_site_intro} or long range ladder~\eqref{eq:double_site}. To conclude that this in fact is the ground state we need to demonstrate the positivity of $H- \mathcal{E}_0$ which as in the isotropic case puts additional constrains on the coupling constants $\{J_n^{\mu}\}$ and $\{K_n^{\mu}\}$.

We consider anisotropic linear exchange models
\begin{align}
    H_n^{1} &= \frac{1}{2}\sum_{j=1}^{L/2} Q_{\alpha_n}(\bfS_{2j-1}, \dots, \bfS_{2j+n-1}) - E_{n, 0}, \nonumber \\
    H_n^{2} &= \frac{1}{2} \sum_{j=1}^{L/2} Q_{\beta_n}(\bfS_{2j}, \dots,\bfS_{2j+n}) - E_{n, 0},
\end{align}
where $Q$ generalizes the isotropic expression to
\begin{equation} \label{aLEM}
    Q_{\gamma}(\bfS_1, \dots, \bfS_n) = \sum_{\mu = x,y,z} \gamma^{\mu} \left(S_1^{\mu} + \dots + S_{n+1}^{\mu} \right)^2,
\end{equation}
with the assumption $\gamma^x + \gamma^y + \gamma^z \geq 0$ and 
\begin{equation}
    E_{n, 0} = \begin{cases}
        \quad 0 , \qquad n \text{ odd}, \\
        - \frac{\mathcal{E}_0}{2}, \qquad n \text{ even}.
    \end{cases}
\end{equation}
are again chosen such that the dimer state (Neel state) has zero eigenvalue for $n$ even (odd).

The relation between coefficients of the anisotropic LEM's and the couplings constants is the component-wise version of~\eqref{eq:alpha_beta_J_Jp}, namely
\begin{equation} \label{alpha_beta_J_K_ani}
    \begin{aligned}
        \alpha_n^{\mu} &= J_n^{\mu} - J_{n+1}^{\mu} - K_{n+1}^{\mu} + K_{n+2}^{\mu}, \\
        \beta_n^{\mu} &= K_n^{\mu}- K_{n+1}^{\mu} - J_{n+1}^{\mu} + J_{n+2}^{\mu}.
    \end{aligned}
\end{equation}

The novelty introduced by the anisotropy is that for a given $n$, as we will soon see, not all the components of $\alpha_n^{\mu}$ or $\beta_n^{\mu}$ must be positive for the corresponding anisotropic LEM to be semi-positive definite. Recall, that in the isotropic case the conditions were simply $\alpha_n, \beta_n \geq 0$. Now the space of possibilities is larger and determined by the spectrum of anisotropic LEMs~\eqref{aLEM}. However, once the semi-positivity of $Q_\gamma(\bfS, \dots, \bfS_n)$ LEMs is understood, the structure of convex cones, described in the previous section, naturally extends to the anisotropic case. 

%\subsection{Anisotropic LEM's and asymmetric quantum tops}

Therefore, the remaining question is to determine when $Q_\gamma(\bfS_1, \dots, \bfS_n)$ is semi-positive definite. It is an operator constructed from the sum of spin operators on different sites and so its spectrum factorises according to the representation theory of $su(2)$:
\begin{equation}
    \frac{1}{2} \otimes \dots \otimes \frac{1}{2} = \begin{cases}
        0 \oplus 1 \oplus \dots \oplus \frac{n+1}{2}, \qquad &n \text{ odd},\\
        \frac{1}{2} \oplus \frac{3}{2} \oplus \dots \oplus \frac{n+1}{2}, \qquad &n \text{ even}.    \end{cases}
\end{equation}
Each block in the direct sum is related to the problem of the asymmetric quantum top for spin-$\mathcal{S}$ operators. The question of semi-positivity of the anisotropic LEM translates then into a question whether the ground-state of the asymmetric quantum top with $\mathcal{S}=1/2$ (or $\mathcal{S}=0$) has the lowest energy among the family of asymmetric quantum tops with $\mathcal{S}=1/2, 3/2, ..., (n+1)/2$ for $n$ even (or $\mathcal{S}=0, 1, ..., (n+1)/2$ for $n$ odd). If this is the case, then $Q_{\gamma}(\bfS_1, \dots, \bfS_n)$ is semi-positive definite. 

To answer this question requires diagonalising the Hamiltonian of the asymmetric quantum top which can be achieved analytically only in special cases and in generality is possible only numerically. 

One special case is the XXZ point where $\gamma^x = \gamma^y=\gamma$ and $\gamma_z \neq \gamma$. In such case the spectrum is 
\begin{equation}
    E(\mathcal{S}, m) = \gamma\, \mathcal{S}(\mathcal{S}+1) + (\gamma_z-\gamma)m^2 ,
\end{equation}
for $m= -\mathcal{S}, -\mathcal{S}+1,  \dots, \mathcal{S},$
with each energy level doubly degenerate. The condition for the semi-positivity of $Q_\gamma$ is then $E(1/2, \pm 1) \leq E((n+1)/2, (n+1)/2)$ which translates into 
\begin{equation} \label{eq:aLEM_XXZ}
    \gamma_z \geq -\frac{2}{n+2} \gamma.
\end{equation}
This condition directly links the magnitude of the anisotropy to the range of the interactions $n$. 

\begin{figure}
   \includegraphics[scale=0.6 ]{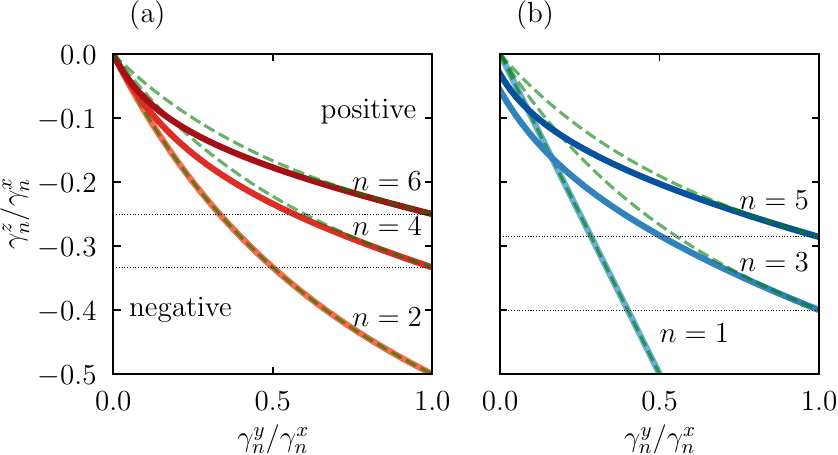}
\caption{The critical lines for the positivity of the anisotropic $Q_\gamma$ operators for even (left panel) and odd (right panel ) ranges $n$ of the couplings. The horizontal lines are the XXZ values of eq.~\eqref{eq:aLEM_XXZ}. For $n=2$ we have an analytical result shown as dashed green lines \cite{Xu2021} which we compare to $n>3$ results from ED in (a) and (b).}
   \label{fig:Q_positivity}
\end{figure}

By numerically diagonalizing the asymmetric quantum top we confirm that the same tendency is in the fully anisotropic case as shown in Fig.~\ref{fig:Q_positivity}. Furthermore, we observe that introducing anisotropy in the $y$-direction requires reducing the anisotropy in the $z$-direction to maintain the semi-positiveness of the anisotropic LEM. A sort of maximal anisotropy principle. 

This principle can be observed analytically for $Q_\gamma(\bfS_1, \bfS_2, \bfS_3)$ because fully anisotropic quantum tops with $n=1,3$ are another special cases where the exact solution can be found. The resulting criterium is
\begin{equation} \label{Q_2_critical}
    \gamma^x \gamma^y + \gamma^y \gamma^z + \gamma^z \gamma^x \geq 0.
\end{equation}
%\milosz{podsumowanie}
Thus, the main finding of this section is that the dimer state is in fact stable under turning on the anisotropy and its stability extends to negative values of the coefficients which is impossible in the isotropic case.  However in the limit of large interaction distance maximal negative anisotropy goes to zero, pointing out to a local-interactions character of this phenomenon.

%For larger values of $n$ the problem can be easily solved numerically through exact diagonalization. In Fig.~\ref{fig:Q_positivity} we show the critical lines 
%milosz{komentarz, ze chcemy by tez $\gamma^x + \gamma^y + \gamma^x \> 0$}

% the ground state energy of the LEMs is known and equal:
% \begin{align}
%     E_{GS}(n)=-n\frac{1}{8}(\gamma^x+\gamma^y+\gamma^z)L,
%     \label{eq:Energy_GS_Aniso_XYZ}
% \end{align}
% where $n=2,4,6...$ are further LEM models with the exact dimer GS.

\begin{figure}
   \includegraphics[scale=0.45]{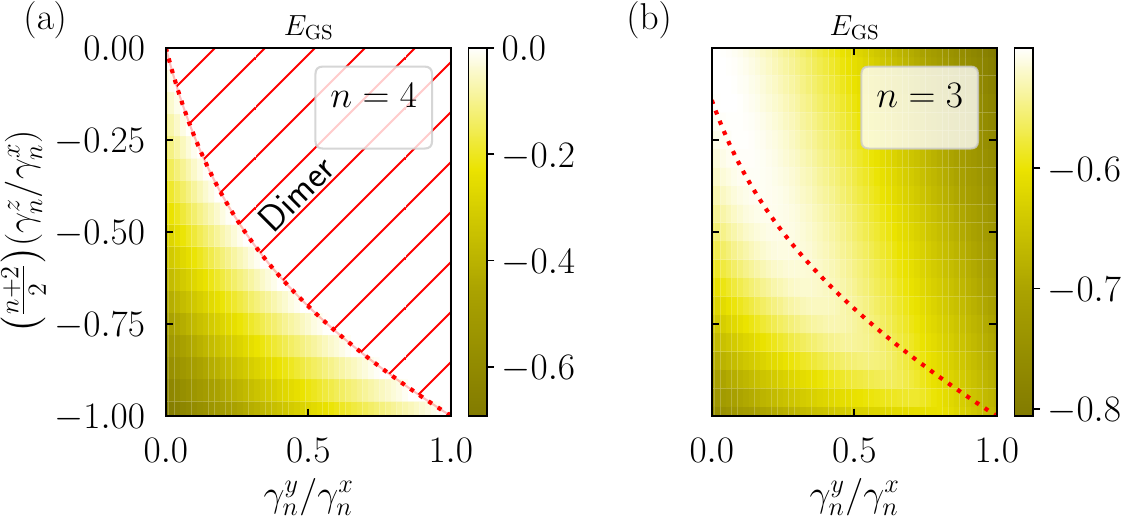}

    \caption{ The ground state energy density of anisotropic LEM Hamiltonians in $\gamma^y$-$\gamma^z$ plane normalised by $\gamma^x$.
    (a) For $n=4$ and above the critical line, the anisotropic LEM is semi-positive definite, the ground state has zero energy, and is the dimer state. Below, the critical line the anisotropic LEM is not positive and the ground state is not anymore the exact dimer.
    (b) For $n=3$ there is an approximate at small anisotropies dimer state we show nonexact dimer phase transition that happens below the red line, where we can see the highest scaled values of energy happen around the phase transition line.
    At $\gamma^y/\gamma^z=1$ the model admits XXZ anisotropy, transitioning to an exact fully polarised state, while for $\gamma^y/\gamma^z<1$ the model is predominantly ferromagnetic, which makes it partially magnetized.}
   \label{fig:DeltazJ2_DeltazJ2p}
\end{figure}

\section{Examples} \label{sec:examples}

We illustrate here the universal construction developed in the previous sections with few examples that capture the essence of the generic situation. First, we will consider systems with short-range interactions and study three models generalizising the Majumdar-Ghosh Hamiltonian in various way. First we will consider its anisotropic version, then we will include interactions of longer-range then $2$ and finally we will consider its realization on the ladder geometry. The last example, will be case of infinite range interactions where we confront the dimerization conditions with the Haldane-Shastry model and deformations thereof. 

We will verify the theoretical construction against numerical exact diagonalizaiton. To probe for the dimer ground state we will use the following observables: ground state energy and dimer entanglement entropy~\cite{DurgaPrasadGoli2013} given by:
\begin{equation}\label{eq:EE_S_dimer}
    \Delta S_{\rm dimer}=\left[S_{\rm vN}(L/2+1)-S_{\rm vN}(L/2)\right]/\ln{(2}),
\end{equation}
where $S_{\rm vN}(x)$ with $x=1, \dots, L$ is the von Neumann entanglement entropy of the subregion $[1, \dots, x]$ of the system. 

In the exact dimer state $\Delta S_{\rm dimer} = 1$ since the entanglement is zero between the bonds of the product state and is maximal and equal to $\ln2$ when computed across the bond formed by the singlet. The same behavior is observed in the triplet product state, however this cannot lead to incorrect determination of the dimer phase. The reason is the following. It requires $J^{II}_n=0$ for the triplet state to be an eigenstate, while the dimer state requires $J^{TT}_n=0$ with $J_n^{II}$ and $J_n^{TT}$ given in eq.~\eqref{eq:dimerisation_pseudospin}. These conditions together imply that $K_n = - J_n$ and, according to eq.~\eqref{eq:alpha_beta_J_Jp}, the Hamiltonian is then not positive-definite. 

If we break the dimerization conditions \eqref{eq:dimerisation_conditions_isotropic} the values of the entropy at the nonexact dimers are different, while in between dimers we find nonzero entropy~\cite{Danu2012}. This provides us with a more precise measure compared to the dimer operator~\cite{white1996dimerization}.

%The construction presented here allows creating much larger family of different models. Below we present only the simplest examples.

%\milosz{do dyskusji w przykladach:} 

\subsection{Anisotropic Majumdar-Ghosh model}

In the notation of anisotropic LEMs, we consider model with $\alpha_1^{\mu} = \beta_1^{\mu}$ and $\alpha_2^{\mu} = \beta_2^{\mu}$ with all higher coefficients equal to zero. This is the most general, two-body Hamiltonian with nearest and next-to-nearest neighbors anisotropic interactions. The dimerization conditions
\begin{equation}
    \alpha_1^{\mu} = 0, \qquad \alpha_2^{x} \alpha_2^{y} + \alpha_2^{y} \alpha_2^{z} + \alpha_2^{z} \alpha_2^{x} \geq 0,
\end{equation}
with the latter being the semi-positivity condition of anisotropic LEM $n=2$ with anisotropies given by $\alpha_2^{\mu}$, as discussed above. In terms of $J_1^{\mu}$ and $J_2^{\mu}$ the two conditions read $J_2^{\mu} = J_1^{\mu}/2$ and (we dropped the index since the inequality is the same for $J_1^{\mu}$ and $J_2^{\mu}$ proportional to each other)
\begin{equation} \label{condition_aMG}
    J^{x} J^{y} + J^{y} J^{z} + J^{z} J^{x} \geq 0.
\end{equation}
We conclude that the anisotropic Majumdar-Ghosh model with the Hamiltonian
\begin{equation}
    H = \sum_{j=1}^L \sum_{\mu = x,y,z}\!\! J^\mu \left(S_j^\mu S_{j+1}^\mu + \frac{1}{2}S_j^\mu S_{j+2}^\mu \right),
\end{equation}
has the dimer phase when the interaction parameters obey~\eqref{condition_aMG}.

The anisotropic Majumdar-Ghosh model was recently studied in~\cite{Xu2021} where exactly the same conditions were found and verified numerically. Here, the same result follows from a direct application of our general formalism.

\subsection{$J_1-J_4$ chain}

We consider now a generalization of the Majumdar-Ghosh chain model ($K_n = J_n$) by introducing interactions of range $3$ and $4$. Thus, we have non-zero $J_1, \dots J_4$ with all higher coefficients equal $0$. In the LEM's construction this corresponds to a linear combination of $H_2$ and $H_4$ models or the non-zero coefficients are $\alpha_2 = \beta_2$ and $\alpha_4 = \beta_4$. We consider both isotropic and anisotropic chains starting with the former.

{\bf Isotropic chain: }
The $J_1-J_2-J_3-J_4$ chain in general has 4 non-zero coupling constants, however recalling the first dimerization condition \eqref{eq:dimerisation_conditions_isotropic}:
we simplify the parameter space by rewriting $J_2$ and $J_4$ in terms of $J_1$ and $J_3$:
\begin{equation}
    J_2=(J_1+J_3)/2,\qquad J_4=J_3/2.
    \label{eq:dimer_iso_trans_sym_rel}
\end{equation}
From the second dimerization condition of~\eqref{eq:dimerisation_conditions_isotropic} we predict to find dimer ground state when,
\begin{equation}
    0  \leq J_3\leq J_1/2,
\end{equation}
which agrees with the earlier results~\cite{nakano1997long}. 
%To get the boundary line we sharpen this inequality and we plot $J_3= J_1/2$ as a red dashed line in Fig.~\ref{fig:n_4_isotropic}(a).\\
We also analyse the necessary condition~\eqref{eq:FM_transition_condition} which describes points in the phase space of the model where the energy of the dimer and ferromangetic states are equal. For the considered model, and taking into account constraints~\eqref{eq:dimer_iso_trans_sym_rel},
the condition is
\begin{align}
    %&J_1+J_2+J_3+J_4=\frac{1}{2}\bigg(3J_1+4J_3\bigg)\geq -\frac{3}{2}J_1,\\
    &J_3\geq -\frac{3}{2}J_1.
\end{align}

\begin{figure}
   \includegraphics[scale=0.5]{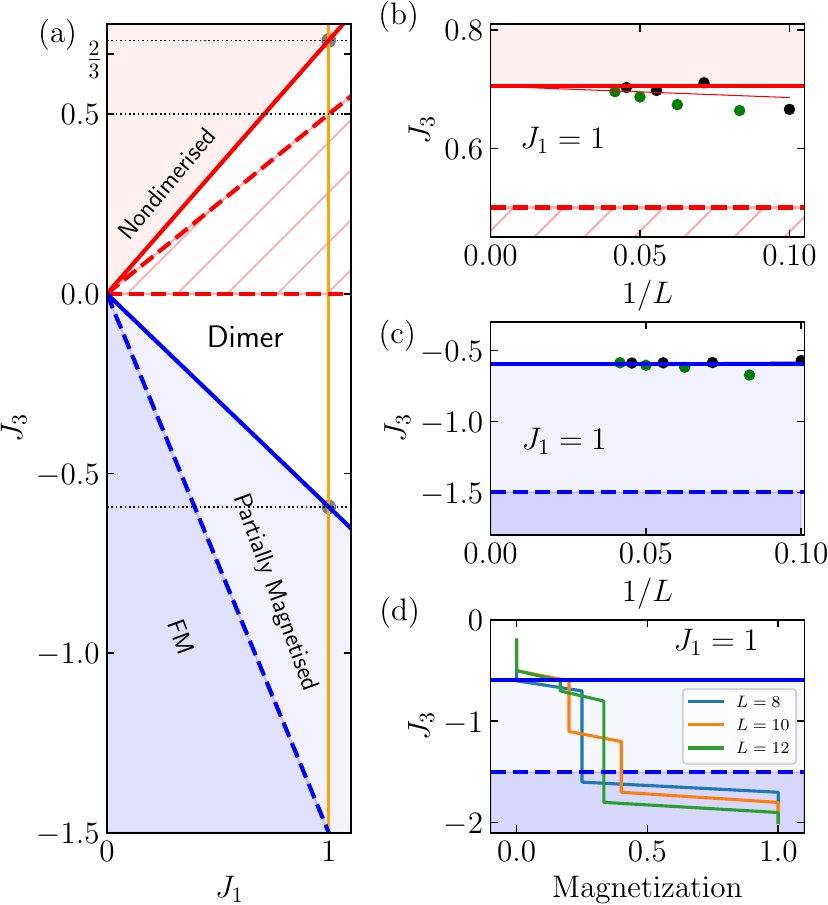}
    \caption{(a) Phase diagram of $J_1-J_2-J_3-J_4$ isotropic chain. The dashed lines represent predicted phase boundaries, while solid lines show phase boundaries deduced from the ground state energy and dimer entropy computed with exact diagonalization in PBC. Finite size scaling results based on the on the ground state energy are shown in (b) and (c).  At positive values we see dimerization up to $J_3\approx2J_1/3$. Meanwhile in the negative values we see that the dimer state disappears at $J_3\approx-J_1/2$ (blue solid line), while polarised state is predicted at $J_3=-3J_1/2$ (blue dashed line).  Magnetization of the whole system for $L=8,10,12$ is shown in (d) and we see that in between dimerized and polarized state the system becomes partially polarized. 
    }
   \label{fig:n_4_isotropic}
\end{figure}

In Fig.~\ref{fig:n_4_isotropic} we present a comparison of the theoretical predictions with the exact diagonalization of the model. We find that the numerical results confirm the theory, namely the dimer phase is found where the theory predicts it and is not found where the theory excludes it. Numerical results suggest that the dimer phase is stable and extends beyond the region where the sufficient conditions are fulfilled. 

Namely, interaction $J_3$ can be both slightly larger then the upper bound $J_3 < J_1/2$ and can also be negative thus going beyond the lower bound. The second case seems especially interesting, since it introduces a competing ferromagnetic interactions into the system. With $J_3$ negative also interactions of range $4$ become ferromagnetic.  

The numerical results seem to suggest that the dimer phase is stable up to around $J_3 = - J_1/2$. At this point $J_2 = J_1/4$ and $J_4 = -J_1/4$ which shows that the value of $J_2$ is much below the Majumdar-Ghosh point and the dimer phase is stabilised by the presence of additional ferromagnetic interactions. %\milosz{cos wiecej o tym mozemy powiedziec?} \jedrzej{cos w stylu:}
Negative values of long range interactions dominate between second neighbour dimers, while the interaction between nearest dimers remains in total positive. This might point out to a hidden long range symmetry protection in a model with negative values of $J_3$. However we might also think in terms of individual bonds, as FM bonds do not compete with AFM bonds for singlet formation, they might be seen as isolating dimers from one another.

When the interactions $J_3$ drop below around $-J_1/2$ the ground state leaves the exact dimer state and enters a partially polarized phase before becoming fully polarized past the $J_3 \approx - 3/2 J_1$, the value where the dimer and the ferromagnetic states have the same energy.

%which we plot as blue dashed line in Fig.~\ref{fig:n_4_isotropic}(a).

{\bf Anisotropic chain:} Allowing for the anisotropic interactions greatly increases the complexity of the problem with now $16$ potentially different coupling strengths. This is partially tamed by $6$ equations generalizing those of eq.~\eqref{eq:dimer_iso_trans_sym_rel} to each of the $3$ spin components. 

To simplify the discussion we focus on the XXZ anisotropy. This further reduces the dimensionality of the phase space to $4$ coupling strengths which is the double of the isotropic case. 
Additionally, the anisotropic LEM's at the XXZ point have analytic solutions which allows to directly infer about their semi-positivity. Indeed, recalling the results of Section~\eqref{sec:anisotropic}, the XXZ LEM $H_{n}^{1,2}$ is semi-positive definite when the coefficients obey
\begin{equation}
    \alpha_n^z \geq - \frac{2}{n+2} \alpha_n,
\end{equation}
where we denote $\alpha_n = \alpha_n^x = \alpha_n^y$. 

The resulting sufficient conditions for the dimer phase~are
\begin{equation}
    \begin{aligned}
    J_2&=(J_1+J_3)/2,\quad &\;J_4=J_3/2, \\
    J_2^z&=(J_1^z+J_3^z)/2,\quad &J_4^z=J_3^z/2,
    \end{aligned}
\end{equation}
and 
%\milosz{(wspolczynniki do sprawdzenia)}\jedrzej{(Sprawdzone, ale jeszcze numerycznie trzeba)}
%\begin{align}
%    J_1^z - 2J_3^z \geq -\frac{2}{3}\left(J_1 - 2J_3\right), \quad J_3^z \geq - \frac{2}{5}J_3,
%\end{align}

\begin{align}
    \frac{1}{2}J_1^z - J_3^z \geq -\frac{2}{4}\left(\frac{1}{2}J_1 - J_3\right), \quad J_3^z \geq - \frac{1}{3}J_3.
\end{align}

As in the isotropic case these conditions completely determine strength of the interactions at ranges $2$ and $4$ and provide bounds on the strengths of the interactions at ranges $1$ and $3$. 

To reduce the dimensionality of the phase space we set $J_3 = J_1/3$ and fix the energy scale such that $J_1 = 1$. This leaves us with two free parameters $J_1^z$ and $J_3^z$ with the sufficient condition for the dimer phase
\begin{equation}\label{eq:J1J2J3J4_aniso_dimer_cond}
    \frac{1}{2}J_1^z -  J_3^z \geq  -\frac{1}{12}, \qquad J_3^z \geq - \frac{1}{9}.
\end{equation}
For transition to FM phase we use \eqref{eq:FM_transition_condition}:
\begin{equation}\label{eq:FM_aniso_chain_transition}
J_3^z\geq-J_1-\frac{1}{2}.
\end{equation} 
\begin{figure}
   \includegraphics[scale=0.45]{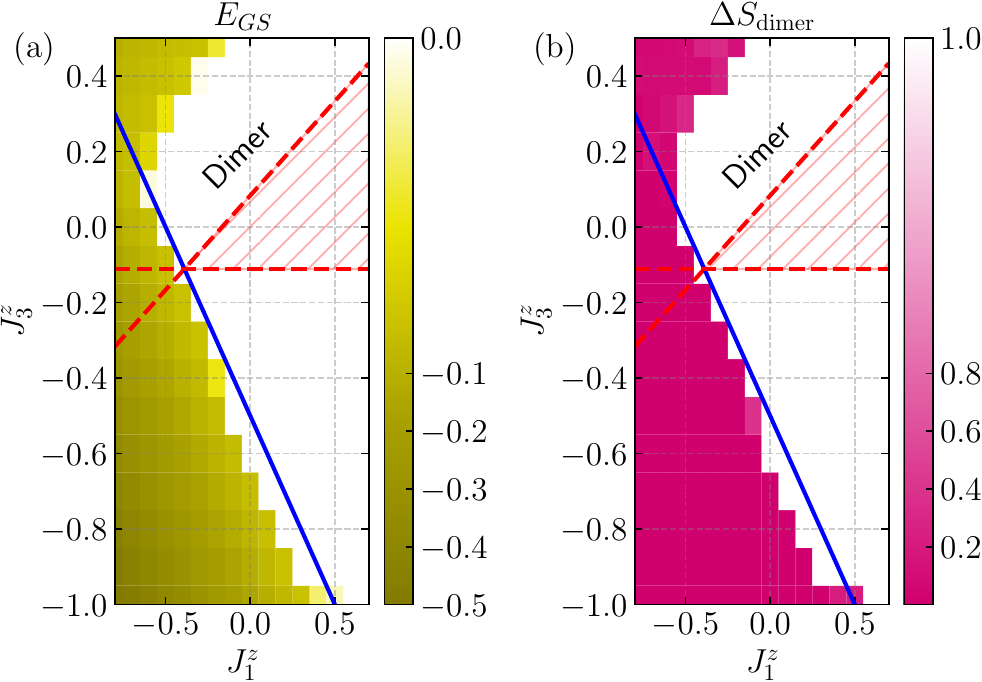}
    \caption{ 
    Phase diagram of $J_1-J_2-J_3-J_4$ anisotropic chain shown by measurements of the ground state energy and dimer entropy done with ED for system of size $L=16$ in OBC. The dimerization conditions predict the dimer phase in the hatched region bounded by dashed red lines. As in the isotropic case we see the conditions are not necessary as we find dimer phase in the whole white region. The blue line denotes phase transition to an exact FM phase. 
    }
   \label{fig:XXZ_n_2_XXZ_n_4_Sdimer_EGS_Plot}
\end{figure}
We plot $E_{\rm GS}$ and $\Delta S_{\rm dimer}$ results from ED computations in Fig.~\ref{fig:XXZ_n_2_XXZ_n_4_Sdimer_EGS_Plot}.
W see that dimer phase is again present in the red lines region predicted by conditions \eqref{eq:J1J2J3J4_aniso_dimer_cond}, however just like in the isotropic case the dimerized region extends beyond that. For $J_3^z<0$ part of the plot the dimer phase stops roughly around the blue line predicted by polarization-dimer energy equality \eqref{eq:FM_aniso_chain_transition} while for $J_3^z>0$ part above the $J_3^z=-\frac{1}{2}J_1^z+\frac{1}{12}$ line.

\subsection{$J_1-K_1-J_2-K_2$ ladder}

We consider here an alternative deformation of the Majumdar-Ghosh model by defining it on the ladder while keeping the interaction to the nearest and next-to-nearest neighbors. 

{\bf Isotropic ladder:} In the isotropic case there are $4$ coupling constants $J_1, J_2, K_1, K_2$. The dimerization conditions~\eqref{eq:dimerisation_conditions_isotropic} put on them the following constraints
\begin{equation}
    \begin{aligned}
        &K_1=J_2+K_2, \\
        &J_1-J_2-K_2\geq 0,\quad J_2\geq 0, \qquad K_2\geq 0.
    \end{aligned} \label{sufficient_J1K1J2K2}
\end{equation}
We also consider the necessary condition~\eqref{eq:FM_transition_condition} which in the present context gives
\begin{equation}
    J_1 \geq - \frac{1}{2} (J_2 + K_2).\label{necessary_J1K1J2K2}
\end{equation}
With $K_1$ fixed, the model depends on $J_1,J_2$ and $K_2$, which we further simplify by measuring the energy in the units of $J_1$ or equivalently setting $J_1=1$. 

\begin{figure}
    \includegraphics[scale=0.5]{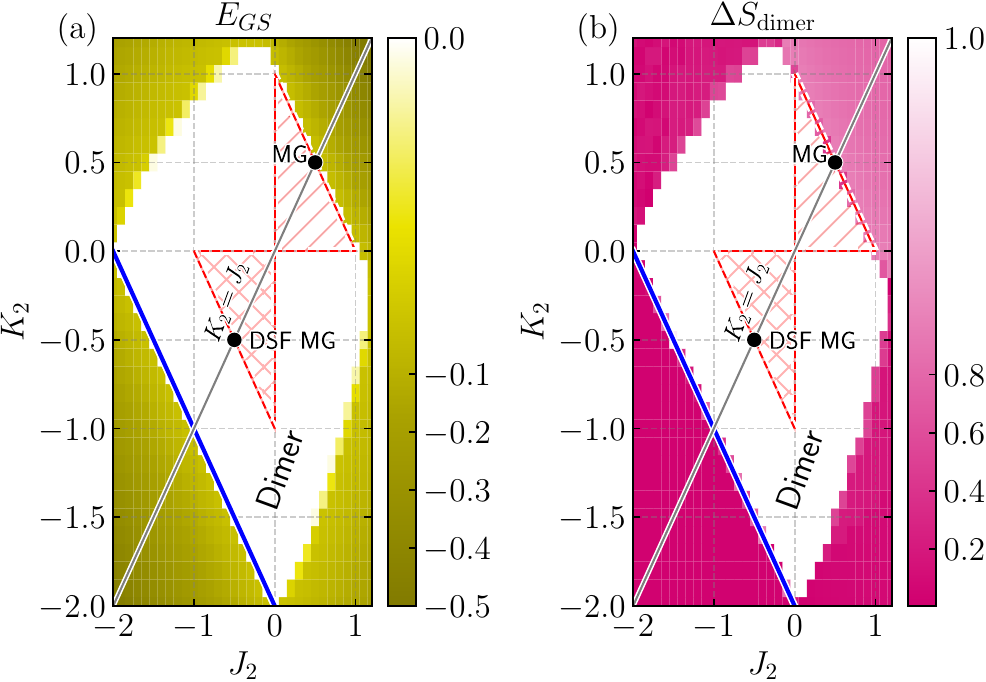}
    \caption{Phase diagrams of $J_1-K_1-J_2-K_2$ isotropic spin chain measured with (a) ground state energy and (b) dimer entropy. We follow the same convention as in Fig.~\ref{fig:XXZ_n_2_XXZ_n_4_Sdimer_EGS_Plot}. 
    %Based on dimerization conditions~\eqref{eq:dimer_iso_trans_sym_rel} we draw the red line $K_2=-J_2+1$ at positive $J_2,K_2$ values with MG at the $J_2=K_2=0.5$ point. On the negative $J_2,K_2$ side we plot phase transition line $K_2=-J_2-2$ to fully polarized state obtained from~\eqref{eq:FM_transition_condition}. 
In addition to the standard region determined by conditions~\eqref{sufficient_J1K1J2K2}, we also denote the complementary region. Within this complementary region, the dimer state is guaranteed by applying a spin-flip operation to the models of the original region. The resulting conditions are~\eqref{spin_flip_J1K1J2K2} Furthermore, we have indicated the Majumdar-Ghosh model and its spin-flipped counterpart. 
     }
   \label{fig:ED_dimer_J1K1J2K2}
\end{figure}

The results of the exact diagonalization are shown in Fig.~\ref{fig:ED_dimer_J1K1J2K2}. We again find the dimer phase inside the region determined by the sufficient conditions~\eqref{sufficient_J1K1J2K2} and we do not find it where the necessary conditions~\eqref{necessary_J1K1J2K2} are not fulfilled. We also find that the dimer phase extends beyond the region of sufficient conditions and in the contrast to the $J_1$--$J_4$ case, it extends all the way to the ferromagnetic line given by the necessary condition~\eqref{necessary_J1K1J2K2}.

Thus, as expected, the imbalance between the rung and leg interactions introduced by $J_1>K_1$ stabilises the dimer phase, as compared with the Majumdar-Ghosh model. What is less expected is the next-to-nearest order interactions can be both ferromagnetic. This happens at the cost of further decreasing $K_1$ and thus increasing the rung-leg imbalance. 

Additionally, the larger dimer phase that the sufficient conditions guarantee can be explained partially explained by invoking the following argument. Let us introduce a double spin flip operation in which pairs of spins $(4j-1, 4j)$ are inverted
\begin{equation}
\uparrow\uparrow\uparrow\uparrow\uparrow\uparrow...\qquad\to\qquad\uparrow\uparrow\downarrow\downarrow\uparrow\uparrow...\, .
\end{equation}
In the ladder geometry with OBC condition this unitary operation has two features. It changes the signs of couplings $J_{4n-2}, J_{4n-1}$ and $K_{4n-3}, K_{4n-2}$ and it leaves the dimer state invariant and thus of the same energy. Therefore, the dimer state remains the ground state of the theory. 

When applied to the studied model, the double spin flip implies changing signs of $K_1$, $J_2$ and $K_2$. Changing appropriately the signs in the dimerization conditions~\eqref{sufficient_J1K1J2K2} leads to 
\begin{equation}
    \begin{aligned}
        &K_1=J_2+K_2, \\
        &J_1+J_2+K_2\geq 0,\quad J_2\leq 0, \qquad K_2\leq 0,
    \end{aligned} \label{spin_flip_J1K1J2K2}
\end{equation}
with the resulting region in the parameter space shown Fig.~\ref{fig:ED_dimer_J1K1J2K2}. Interestingly, the spin-flipped version of the Majumdar-Ghosh model (with $K_1 = - J_1$ and $J_2 = K_2 = - J_1/2$) is located roughly in the center of the dimer phase and therefore is more stable then the standard MG model.

% We consider three cases: 
% \begin{enumerate}
%     \item $J_1\geq J_2,K_2>0,J_1\geq K_1>0$,
%     \item $J_2<0,K_2,K_1<0$,
%     \item $J_1\geq J_2>0,K_2<0$ or $J_2<0,J_1\geq K_2>0$.
% \end{enumerate}

% To find phase transition line in the first case we just look at the equality, $J_1=J_2+K_2$ getting equation:
% \begin{equation}
%     K_2=-J_2+J_1\label{eq:J1_K1_phase_trans_nondimer}
% \end{equation}

% For the second case we got a phase transition line to a polarized state, which we compare to energy of the dimer state:
% \begin{align}
%     &\frac{E_{\rm FM}}{L}=\frac{1}{8}(J_1+K_1+J_2+K_2)=\frac{1}{8}(1+2(J_2+K_2)),\\
%     &\frac{E_{\rm dimer}}{L}=-\frac{3}{8}J_1=-\frac{3}{8},\\
%     &K_2=-J_2-2 \label{eq:J1-K1_phase_polarised}
% \end{align}

% For the third case of mixed signs of $J_2,K_2$ we are not allowed to impose such conditions based only on the relations or energy. We only impose boundaries on positive interaction $J_1\geq J_2$ (or $J_1\geq K_2$).

% To check 

%  to investigate numerically 1D cuts of the diagram for fixed $J_2$, as a function of $K_2$.

% From the figure we see that the predicted phase transitions for cases I and II are in agreement with results from ED, while additionally the boundaries in the quadrants where $J_2>0,K_2<0$ approximately scale linearly from the critical points from $(J_2,K_2)=0,-2J_1$ and   $(J_2,K_2)=J_1,0$ and symmetrically in the 
% $J_2<0,K_2>0$ quadrant.

% We can write the Hamiltonian of the ladder as a sum of three LEMs, parameterized by $\alpha_1,\alpha_2,\beta_2$:

% \begin{align}
% &H=\alpha_1H_1^1+\alpha_2H^1_2+\beta_2H^2_2
% \label{eq:Ladder_iso_boundaries_relations}
% \end{align}
% We find the above representation for Hamiltonians of the systems associated with predicted phase transition lines by representing $\alpha,\beta$ coefficients with $J,K$ using equations \eqref{eq:alpha_beta_J_Jp}.
% For the transition to fully polarized in $J_2<0,K_2<0$ region \eqref{eq:J1-K1_phase_polarised} , we have:
% \begin{align}
% &\alpha_1=3J_1-2J_2\qquad\alpha_2=-\beta_2=2J_1
% \end{align}
% For the transition out of dimer state in $J_2>0,K_2>0$ region \eqref{eq:J1_K1_phase_trans_nondimer} we have:
% \begin{align}
% &\alpha_1=0\qquad\alpha_2=-\beta_2=2J_2-J_1
% \end{align}.

%\subsection{Anisotropic $J_1-K_1-J_2-K_2$ chain(ladder)}
{\bf Anisotropic ladder:} Previous results might have given an impression that the dimer phase is common in long-range spin and ladder Hamiltonians. While this is true to some extent, in the same time, the dimer phase requires certain fine tuning. This tuning appears through the conditions $\beta_{2n-1}=0$ which so far in the numerical studies were always satisfied. To exhibit their role we introduce the anisotropy to the Majumdar-Ghosh ladder. 

The condition $\beta_{2n-1} = 0$, in either isotropic or aniotropic model is very strong, as it guarantees that the dimer state is an eigenstate of the model. If it is not fulfilled the dimer state is likely not the eigenstate and hence the ground state of the model cannot be a dimer. The generalization of the first condition of~\eqref{sufficient_J1K1J2K2} to the anisotropic case is
\begin{equation}
    K_1 = J_2 + K_2, \qquad K_1^z = J_2^z + K_2^z,
\end{equation}
where we again allowed for the anisotropy only in the $z$-direction. The consequence of these relations is that anisotropy in the next-to-nearest interactions equals the anisotropy in the nearest neighbor interactions. In the following we will test it by assuming that the nearest neighbor interactions are isotropic, $K_1^z = K_1$. Furthermore, we assume that $K_1=J_1$. The resulting model is $J_1$--$J_2$--$K_2$ with anisotropy $J_2^z$ and $K_2^z$ and the prediction is that the dimer phase is only when
\begin{equation} \label{ladder_anisotropic_condition}
    J_1 = J_2^z + K_2^z.
\end{equation}

\begin{figure}
   \includegraphics[scale=0.55]{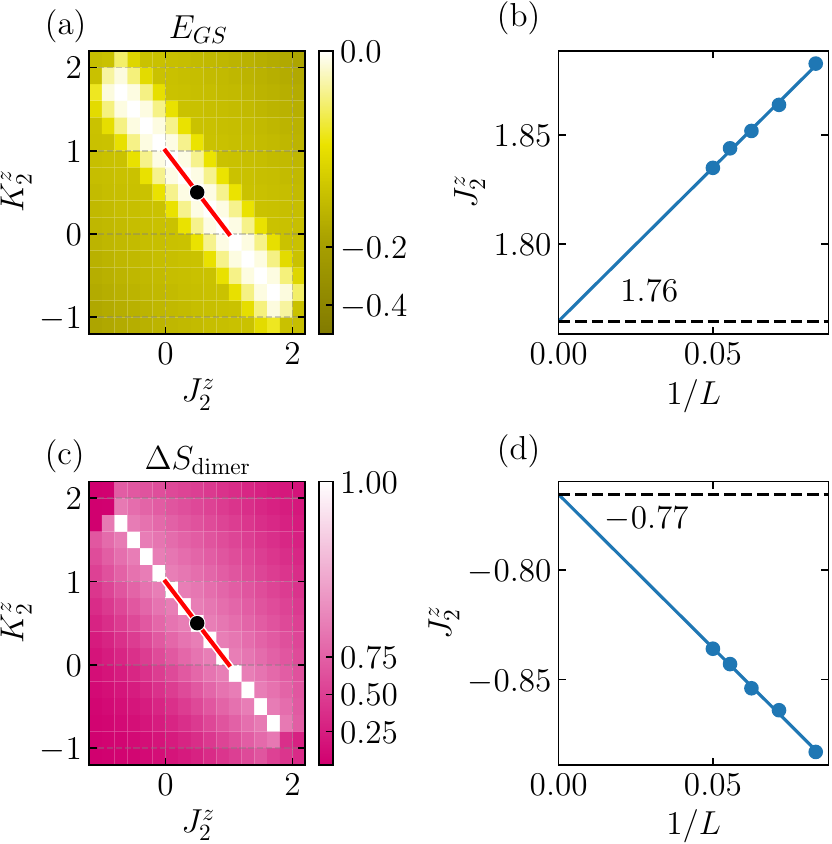}
    \caption{Plots illustrating dimerization region of $J_1-K_1-J_2-K_2$ anisotropic ladder (a) in ground state energy $E_{\rm GS}$  (b) Dimer entropy $\Delta S_{\rm dimer}$.
    The dimerization conditions (red line) lies in the white region where the dimer ground state is found. The edges of this region extend again beyond the theoretical predictions with values obtained by finite size scaling for $J_2^z>0$ in (c) as $J^{z,c}\approx 1.76$ and for $J_2^z<0$ in (d) as $J^{z,c}_2\approx -0.77$.
    }
   \label{fig:ED_Energy_GS_J1J2K2_aniso}
\end{figure}

The results of the exact diagonalization are shown in Fig.~\ref{fig:ED_Energy_GS_J1J2K2_aniso}. They clearly illustrate that the dimer phase is only along the line described by eq.~\eqref{ladder_anisotropic_condition}. However, as before we see the $J_2^z,K^z_2$ values are not necessarily positive (red line), transitioning out of dimer state in finite size scaling limit of $J^z_2\approx1.77$ for positive $J^z_2$ and $J^z_2\approx-0.77$ for negative $J^z_2$. 

\subsection{Infinite interactions range}

As a last example, we consider a spin chain with all-to-all interactions. The paradigmatic model of such situation is the Haldane-Shastry chain~\cite{PhysRevLett.60.635,PhysRevLett.60.639}. Here we consider its generalization, known as Inozemtsev chain, with interpolates between the Haldane-Shastry potential and short-range XXX Hamiltonian~~\cite{1990JSP....59.1143I,Inozemtsev2000,Klabbers_Lamers_Inozemtsev}. The Hamiltonian of the Inozemtsev chain is
\begin{equation} \label{HS}
    H_{\rm Ino} = \frac{1}{2}\sum_{j=1}^L \sum_{n=1}^{L-1} J_n^{\rm Ino}\bfS_j \cdot \bfS_{j+n}.
 \end{equation}
Following~\cite{Klabbers_Lamers_Inozemtsev}, the coupling coefficients are
\begin{equation}
    J_n^{\rm Ino} = \frac{\sinh^2 \kappa}{\kappa^2} \left( \wp(n) + \frac{\eta_2}{\omega}\right),
\end{equation}
and are determined by the Weierstrass elliptic function $\wp(n)$ and parameter $\kappa$. The Weierstrass elliptic function is characterised by two periods which in the present context are $(N, \omega)$ with $\omega = i\pi/\kappa$. Additionally, constant  $\eta_2 = \zeta(\omega)$ where $\zeta(z)$ is the Riemann zeta function with the same two periods $(N, \omega)$. 

Parameter $\kappa$ controls the range of the interactions: $\kappa \rightarrow \infty$ yields nearest neighbor XXX spin chain, while $\lim \kappa \rightarrow 0$ corresponds to the Haldane-Shastry model with couplings
\begin{equation}
    J_n^{\rm HS} = \frac{1}{\left(\sin \pi n /N\right)^2}.
\end{equation}

With the help of the periodicity of the coefficients, $J_n^{\rm Iso} = J_{N-n}^{\rm Iso}$, the Inozemtsev Hamiltonian~\eqref{HS} takes the form of the long-range chain Hamiltonian~\eqref{eq:single_site_intro} with $J_n = (J_{n}^{\rm Ino} + J_{N-n}^{\rm Ino})/2$ for $n = 1, \dots, n_{\rm max}$ with $n_{\rm max} = N/2$. The coupling constants obey the convexity property but not the self-averaging relation, c.f. eq~\eqref{eq:dimerisation_conditions_isotropic}. Therefore, the dimer state is not the ground state of the Inozemtsev nor of the Haldane-Shastry chains. In fact it is not even an eigenstate of the theory. This is in agreement with the known results for their ground-states~\cite{Haldane1988,Inozemtsev1996}

%which for the Haldane-Shastry is the Guzwiller-Jastrow resonating valence bound state~\cite{Haldane1988} and for Inozemtsev chain is not of the product form~\cite{Inozemtsev1996}.

\begin{figure}
    \includegraphics[scale=0.4]{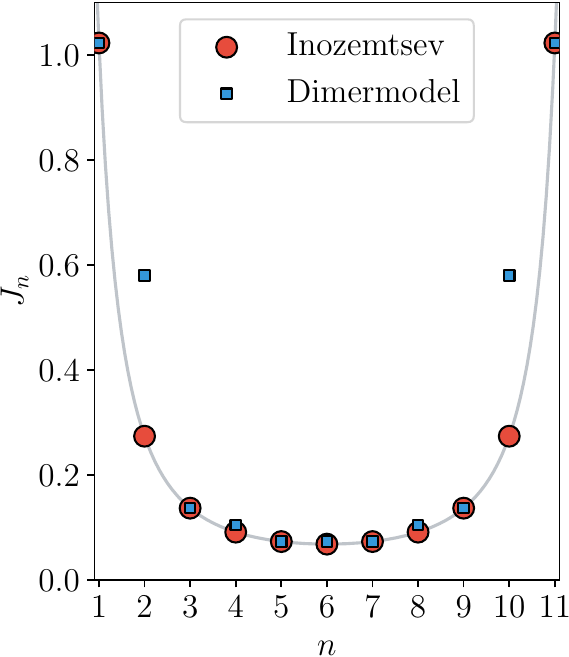}
    \caption{The coupling constants in the Inozemtsev chain of length $L=12$ and in the corresponding dimer chain with odd coupling constants self-averaged. The latter chain has the exact dimer ground state.}
    \label{fig:Inozemtsev}
\end{figure}
%\milosz{cytat do pracy gdzie robia to samo dla HS}
%\jedrzej{To chyba ten}
However, we can easily come up with a deformation of the Inozemtsev model which has the exact dimer states as its ground state. This possibility was first observed for the Caloger-Sutherland case in~~\cite{nakano1997long}.  Namely, for odd $n$ we keep the original couplings from the model while the even ones are specified by the self-averaging property. The resulting coupling coefficients are shown in Fig.~\ref{fig:Inozemtsev} together with a comparison to the Inozemtsev values.

\section{Conclusions and Discussion}~\label{sec:final}

In this study, we established a general framework determining sufficient conditions for existence of the dimer ground state in a wide class of long-range spin chains Hamiltonians with bond alteration and anisotropic interactions. In many-dimensional spaces spanned by the model interactions, the dimer states are guaranteed to exist in convex polytopes determined by the rules of dimerization. These theoretical predictions were verified for the Majumdar-Ghosh model and its generalizations revealing the existence of the dimer ground-state with aniostropic and/or long-range interactions and/or bond alterations. The result of this work provide thus exact reference points for further studies of these classes of models. The results are non-perturbative and valid in finite systems.

The dimerization conditions, when additional symmetries of a Hamiltonian are present, can serve also as seeds in the phase space from which other regions of the exact dimer ground state can be determined. We illustrated this in the example of the Majumdar-Ghosh model on the ladder. 

It is also worth highlighting one partial result that is the exact bound on the spectrum. Namely, for the coefficients $\{J_n\}$ and $\{K_n\}$ such that $\alpha_n \geq 0$ and $\beta_n \geq 0$, the spectrum of the Hamiltonians~\eqref{eq:double_site} is bounded from below by ${\rm min}\left({\rm spec} (H)\right) \geq E_0$ with $E_0$ given in eq.~\eqref{LEM_shift}. This is an exact bound valid for large class of models and independent of the structure of their ground state. 

The framework introduced here can be generalized in various ways. Possible directions include considering systems with $3$-spins or larger interactions~\cite{Takano1994,Michaud2012,Michaud2013} or higher spin models~\cite{nakano1997long,matera2014phase,Lamas2015,Hikihara2017,Petrovich2024}. Another direction is to change the target ground state from the dimer to either generalized dimers~\cite{Petrovich2024} or finitely correlated states \cite{Fannes1992,Michalakis2006,Fanizza2023,Roon2025,Affleck1987,affleck1988valence,Petrovich2022,aizenman1994geometric,Chen2012,Zhu2024}.

The results of our work also raise the question of the excitation gap. Characterizing it provides deeper insight into the stability of the dimer phase across different coupling strengths and under an external magnetic field. While the standard Majumdar-Ghosh model and its immediate generalizations are gapped~\cite{takano1994excitation,Danu2012}, modifications like ferromagnetic interactions, ladders, or anisotropies could alter the excited state spectrum. For isotropic interactions, the numerical results confirm that the dimer ground state is gapped in the region where the sufficient conditions hold. Further studies in this direction are left for future work.

\begin{acknowledgments}
We acknowledge inspiring discussions and feedbacks on this and related topics from Fabien Alet, Piotr Chudzinski, Fabian Essler and Jacek Herbrych.
The authors acknowledge support by the National Science Centre (NCN), Poland via project 2022/47/B/ST2/03334.
\end{acknowledgments}

%\section*{Data Availability}

\vspace{0.25cm}

\noindent {\bf Data Availability:} All diagrams, plots, underlying data, and the Python scripts used for computation and plotting are available online~\cite{zenodo2026}.

\appendix

\section{Pseudospin representation} \label{app:exact_dimers}

This appendix outlines a few intermediate steps in rewriting the long-range ladder Hamiltonian using the pseudospin representation. 

We start by writing the long-range ladder Hamiltonian~\eqref{eq:double_site} as a sum of fixed-range Hamiltonians distinguishing the even and odd range
\begin{equation}
    H = \sum_{k \geq 1} \left(H_{2k-1} + H_{2k}\right),
\end{equation}
where
\begin{equation} \label{fixed_range_ladder}
    H_k = \sum_{j=1}^{L/2} \left(J_k \bfS_{2j-1} \cdot \bfS_{2j+k-1} + J_k'\bfS_{2j} \cdot \bfS_{2j+k}\right).
\end{equation}
In the next step we rewrite $H_k$ in terms of the pseudospin operators. 
The inverse transformation to~\eqref{pseudospin} mapping physical spin operators into pseudo-spin operators is
\begin{equation}
    \bfS_{2i} = \frac{{\bfI}_i + {\bfK}_i}{2}, \qquad \bfS_{2i-1} = \frac{{\bfI}_i - {\bfK}_i}{2}.
    \label{eq:pseudo_to_spin}
\end{equation}For even range
\begin{align}
    H_{2k} = \frac{1}{4}\sum_{j=1}^{L/2} &\left[ \left(J_{2k}' + J_{2k}\right)\left(\bfI_j \cdot \bfI_{j+k} + \bfK_j \cdot \bfK_{j+k}\right) \right. \nonumber \\
    &\left. \left(J_{2k}' - J_{2k}\right)\left(\bfI_j \cdot \bfK_{j+k} + \bfK_j \cdot \bfI_{j+k}\right)\right],
\end{align}
while for odd range 
%\begin{align}
%    H_{2k-1} =& \frac{1}{4} \sum_{j=1}^{N/2} \left[ J_{2k-1}\left({\bf I}_j\cdot {\bf I}_{j+k-1} + \bfI_i \cdot \bfK_{j+k-1} - \bfK_i \cdot \bfI_{j+k-1} \right.\right. \nonumber \\
%    &\left.\left. - {\bf K}_j\cdot {\bf K}_{j+k-1}\right) + J_{2k-1}'\left(\bfI_j \cdot \bfI_{j+k} - \bfI_i \cdot \bfK_{j+k} \right.\right.  \nonumber \\
%    &\left.\left. + \bfK_j \cdot \bfI_{j+k} - \bfK_j \cdot \bfK_{j+k}\right)\right].
%\end{align}
\begin{align}
    H_{2k+1} =& \frac{1}{4} \sum_{j=1}^{L/2} \left[ J_{2k+1}\left(\bfI_j - \bfK_j\right) \cdot \left(\bfI_{j+k} + \bfK_{j+k}\right) \right. \nonumber \\
    &+ \left. J_{2k+1}'\left(\bfI_j + \bfK_j\right) \cdot \left(\bfI_{j+k+1} - \bfK_{j+k+1}\right)\right].
\end{align}
After summing over $k$ and rearranging the terms we obtain
\begin{equation} \label{H_PS_plus_identity}
    H = H_{\rm PS} + J_1 H_{XXX}^{\rm rungs},
\end{equation}
with
\begin{align} 
    H_{\rm PS} = \frac{1}{4} \sum_{j=1}^{L/2} \sum_{k\geq 1} \left( J_k^{II} \bfI_j \cdot \bfI_{i+k} + J_k^{KK} \bfK_j \cdot \bfK_{j+k} \right. \nonumber \\
    \left. + J_k^{IK} \bfI_j \cdot \bfK_{j+k} + J_k^{KI} \bfK_j \cdot \bfI_{j+k}  \right), \label{H_PS}
\end{align} 
and the "XXX Hamiltonian" acting on the rungs of the ladder,
\begin{equation}
    H_{XXX}^{\rm rungs} = \sum_{j=1}^{L/2} \bfS_{2j-1}\cdot \bfS_{2j}.
\end{equation}
The coefficients of the $H_{PS}$ Hamiltonian are reported in the main text in eq.~\eqref{eq:dimerisation_pseudospin}.
The above method is taken from Lamas and Matera works \cite{matera2014phase,Lamas2015}.

The case of anisotropic interactions is similar.  Each component’s computations can be performed independently and the resulting relations are identical for each. The dimerization condition is in eq.~\eqref{dimerization_condition_ani} of the main text.

% Let's separate the $H_{\rm PS}$ Hamiltonian into different coefficients:
% \begin{align} 
%     &H_{\rm PS} = \frac{1}{4} \sum_{j=1}^{N/2} \sum_{k\geq 1} \left( J_k^{II} \bfI_j \cdot \bfI_{j+k} + J_k^{KK} \bfK_j \cdot \bfK_{j+k} \right. \nonumber \\
%     &\left. + J_k^{IK} \bfI_j \cdot \bfK_{j+k} + J_k^{KI} \bfK_j \cdot \bfI_{j+k}  \right)
% \end{align} 
% We know that with the above Hamiltonian we get the interactions of the form:
% \begin{align}
%     &\bfI_j \cdot \bfI_{j+k}=\sum_{\mu}I^{\mu}_{j}I^{\mu}_{j+k}=\\
%     &=\sum_{\mu}(S_{2j}^\mu S_{2(j+k)+2}^\mu+S_{2j-1}^\mu S_{2(j+k)+2}^\mu+\\
%     &+(S_{2j}^\mu S_{2(j+k)+1}^\mu+S_{2j-1}^\mu S_{2(j+k)+2}^\mu))
% \end{align}
% As this is separable into $x,y,z$ components we obtain independent equations for ($\bfI_j \cdot \bfI_{j+k}$,$\bfK_j \cdot \bfK_{j+k}$,$\bfI_j \cdot \bfK_{j+k}$,$\bfK_j \cdot \bfI_{j+k}$,):
% \begin{equation}
% \begin{aligned}
%     J_{k,\mu}^{II} &= J_{2k,\mu}' + J_{2k,\mu} + J_{2k-1,\mu}' + J_{2k+1,\mu}, \\
%     J_{k,\mu}^{IK} &= J_{2k,\mu}' - J_{2k,\mu} - J_{2k-1,\mu}' + J_{2k+1,\mu}, \\
%     J_{k,\mu}^{KI} &= J_{2k,\mu}' - J_{2k,\mu} + J_{2k-1,\mu}' - J_{2k+1,\mu}, \\
%     J_{k,\mu}^{KK} &= J_{2k,\mu}' + J_{2k,\mu} - J_{2k-1,\mu}' - J_{2k+1,\mu}.
% \end{aligned}\label{eq:dim_cond_XYZ}
% \end{equation}

\section{Semi-positivity of the linear exchange models} \label{app:LEM}

We show here that linear exchange models $H_n^1$ and $H_n^2$ defined in~\eqref{LEM_def} are semi-positive definite. For $n$ odd this is evident from the definition. For $n$ even we will show that the spectrum is still positive after subtracting the constant $\mathcal{E}_0/2$. 

The proof is a slight generalization of the results of~\cite{Kumar2002} to the ladder case. There it was shown that the spectrum of the Hamiltonian defined on a chain (without the bond alteration)
\begin{equation}
    H[\mathfrak{B}_{2\nu+1}] = \sum_{i=1}^L \sum_{j=1}^{2\nu} \left(2\nu + 1 - j\right)\bfS_i \bfS_{i+j},
\end{equation}
is bounded from below by $- 2\nu \mathcal{E}_0$ with the dimer state saturating the bound. This Hamiltonian can be presented as
\begin{equation}
    H[\mathfrak{B}_{2\nu+1}] = H_{2\nu} - 2\nu \mathcal{E}_0.
\end{equation}
with
\begin{equation} \label{app_LEM}
    H_n = \frac{1}{2}\sum_{j=1}^L \left(S_j + S_{j+1} + \dots S_{j+n} \right)^2 - \mathcal{E}_0.
\end{equation}
This shows that $H_n$ is semi-positive for $n$ even.

We observe that $H_n = H_n^1 + H_n^2$. Therefore, from the semipositivity of $H_n$ we would like to infer the semi-positivity of $H_n^{1,2}$. 

This is the case because the LEM Hamiltonians~\eqref{app_LEM} are frustration free. This implies that the total Hamiltonian and local Hamiltonians 
\begin{equation}
    h_n^{(j)} = \left(S_j + S_{j+1} + \dots S_{j+n} \right)^2,
\end{equation}
have the same ground state. 

Therefore, when the Hamiltonian contains exactly every second local hamiltonian $h_n^{(j)}$ then the ground state energy is the half of the ground state energy of the full model. This shows that the Hamiltonians $H_n^{1,2}$, as defined in~\eqref{LEM_def} are semi-positive definite.

\section{Asymmetric quantum top} \label{app:top}

Asymmetric quantum top is defined by the following Hamiltonian
\begin{equation}
    H_{\rm top} = A \mathcal{S}_x^2 + B \mathcal{S}_y^2 + C \mathcal{S}_z^2,
\end{equation}
where $\mathcal{S}_\mu$ are spin-$S$ operators and we assume $C \leq B \leq A$. To analyse the spectrum of the quantum top we introduce the spin raising and lowering operators $\mathcal{S}_{\pm} = \mathcal{S}_x \pm i \mathcal{S}_y$. In terms of them, the Hamiltonian is
\begin{equation}
    H_{\rm top} = \frac{A - B}{4} \left(\mathcal{S}_+^2 + \mathcal{S}_-^2\right) + C\mathcal{S}_z^2 + \frac{A+B}{2}\left(\mathbf{\mathcal{S}}^2 - \mathcal{S}_z^2\right),
\end{equation}
and is tridiagonal in the basis $|S, m\rangle$ where $S$ is the total spin and $m= -S, -S + 1,\dots, S$ is its projection on the $z$ axis. In the partially symmetric case $A=B$, the Hamiltonian is diagonal and its spectrum can be easily found. The more general case requires numerical solutions beyond the small $S$ cases that can be solved analytically.

The non-zero matrix elements of $H_{\rm top}$ in the basis $|S, m\rangle$ are
\begin{equation}
    \begin{aligned}
        \langle S, m|\mathcal{S}_z^2|S, m\rangle &= m^2, \\
        \langle S, m|\mathcal{S}^2|S, m\rangle &= S(S+1), \\
        \langle S, m\pm 2|\mathcal{S}_\pm|S, m\rangle &= C_{S, m} C_{S, m\pm 1},
    \end{aligned}
\end{equation}
where
\begin{equation}
    C_{S, m} = \sqrt{\left(S(S+1) - m(m\pm 1)\right)}.
\end{equation}
The resulting eigenvalue problem can be solved explicitly for small $S$. The spectra are
\begin{equation}
    \begin{aligned}
        {\rm spec}(S=1/2) &= \left\{\frac{1}{4}\left(A+B+C\right)\right\}, \\
        {\rm spec}(S=1) &= \{A+B,\, A+C,\, B+C\}, \\
        {\rm spec}(S=3/2) &= \left\{\frac{5}{4}\left(A+B+C\right) \pm \sqrt{D}\right\},
    \end{aligned}
\end{equation}
where
\begin{equation}
    D = A^2 + B^2 + C^2 - AB - BC - CA.
\end{equation}
For $S=1/2$ and $S=3/2$ the spectra are doubly degenerate. For higher $S$ spectra can be easily found numerically.

In the application to the asymetric LEMs, we are interested in situations when spectrum for $S=1/2$ lies below the spectrum for $S=3/2$. Comparing the two spectra we find that the condition~is
\begin{equation}
    AB + BC + CA \geq 0.
\end{equation}
This condition, with the notation adjusted to the LEM Hamiltonian is given in eq.~\eqref{Q_2_critical}.

The XXZ point where $A=B$ allows for an exact determination of the spectra for any $S$,
\begin{equation}
    E(S,m ) = A S(S+1)+ (C-A)m^2, \qquad A=B.
\end{equation}

\section{Triplet states from rotation in isotropic models} \label{app:triplets}

We can generalize the problem from dimer (singlet) product state to triplet product state. 
%As mentioned in App.~\ref{app:LEM}, the model can be constructed with projection operators to also allow for non-polarised triplet valence bond states by changing parameter $q$ from $+1$ to $-1$, preserving the product state structure\eqref{product_state}. 
To map the Hamiltonian from the singlet to the triplet ground state, we introduce local unitary $\pi$-rotation operator
\begin{align}
    &U_j=\exp{\bigg(i\pi S^z_j\bigg)},\quad U_j^\dagger S^z_jU_j=S^z_j,\\
    &U_j^\dagger S^\mu_jU_j=(-1)^{(j+1)}S^\mu_j,~~\mu=x,y.
\end{align}
Acting with them on the Hamiltonian changes only the sign in odd distance $n$ interactions:
\begin{align}
    &U_j^\dagger(S_j^xS_{j+n}^x+S_j^yS_{j+n}^y+S_j^zS_{j+n}^z)U_j\\
    &=-(S_j^xS_{j+n}^x+S_j^yS_{j+n}^y)+S_j^zS_{j+n}^z.
\end{align}
We apply this unitary operator to the Hamiltonian of the $J_1-K_1-J_2-K_2$ model. 

\begin{figure}[ht!]
   \includegraphics[scale=0.6]{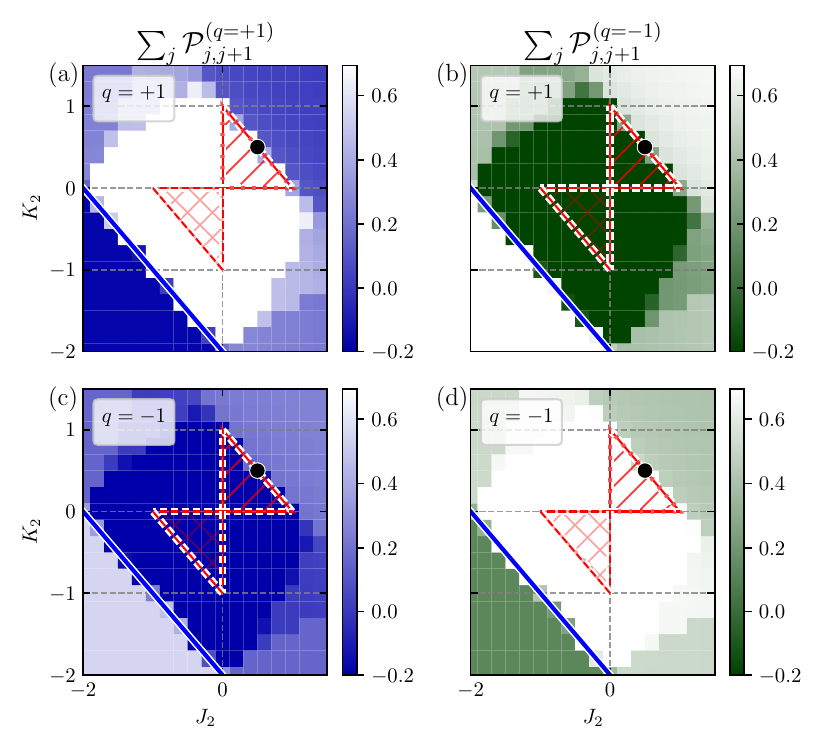}
    \caption{
    Plots of sums of singlet (triplet) projection operator in right (left) column.
    We present results for $q=+1$ in a) and b) and for $q=-1$ in c) and d). We see that while the dimer state region does not change its shape, the projector operators measurements show the change of building blocks from singlets to triplets in product state ground state.
    }
   \label{fig:XXZ_n_2_XXZ_n_4_proj_Plot}
\end{figure}

To see if the dimer structure changed from singlets to triplets, we perform measurements with projection operators on the respective ground state. These can be constructed from the Temperley-Lieb operators $e_i=(q+q^{-1})\mathcal{P}^{(q)}_{j,j+1}=d\mathcal{P}^{(q)}_{j,j+1}$~\cite{Nichols2006}:
\begin{align}
    &\mathcal{P}_{j,j+1}^{(q)}=\frac{1}{d}e_j=\\
    &\bigg((-\frac{2}{d})(S_j^xS_{j+1}^x+S_j^yS_{j+1}^y)-S_j^zS_{j+1}^z+\frac{1}{4}\nonumber\\
    &-\frac{(q-q^{-1})}{2d}(S_j^z-S_{j+1}^z)\bigg).\nonumber
\end{align}
At $q=1$ ($q=-1$) with $d=2$ ($d=-2$) we get the projection operator onto a valence bond singlet (triplet):
\begin{align}\label{eq:projection_singlet_triplet}
    \mathcal{P}_{j,j+1}^{(q=\pm1)}=\mp\big(S_j^xS_{j+1}^x+S_j^yS_{j+1}^y\big)-S_j^zS_{j+1}^z+\frac{1}{4}.
\end{align} 
The results are presented in Fig.~\ref{fig:XXZ_n_2_XXZ_n_4_proj_Plot}. We show them sums of singlet $\mathcal{P}^{(q=+1)}$ or triplet $\mathcal{P}^{(q=-1)}$ projection operators measurements along the whole chain, divided by the length of the system.
We see in Fig.~\ref{fig:XXZ_n_2_XXZ_n_4_proj_Plot}(a) that the region with singlet states agrees with results from energy and dimer entropy from Fig.~\ref{fig:ED_dimer_J1K1J2K2}, and does not feature any triplet states as seen in Fig.~\ref{fig:XXZ_n_2_XXZ_n_4_proj_Plot}(b).
After flipping $q=1\to q=-1$ the singlet projection return negative results in Fig.~\ref{fig:XXZ_n_2_XXZ_n_4_proj_Plot}(c), while now we get nonzero values from triplet projections in Fig.~\ref{fig:XXZ_n_2_XXZ_n_4_proj_Plot}(d). Thus numerical results confirm the structure discussed in App.~\ref{app:LEM}.

\bibliography{refs}
\end{document}